\documentclass[fleqn,usenatbib]{mnras}

\usepackage[T1]{fontenc}
\usepackage{ae,aecompl}
\usepackage{newtxtext,newtxmath}
\usepackage{txfonts}
\usepackage{mathptmx}
\usepackage{graphicx}
\usepackage{amssymb}
\usepackage{color}

\DeclareMathAlphabet{\pazocal}{OMS}{zplm}{m}{n}

\newcommand{\comment}[1]{} 
\newcommand*{\satellite}[1]{\textit{#1}}

\newcommand*{\spitzer}{\satellite{Spitzer}}

\title[AGN clustering at 0.5 < z < 4.5]{The clustering of X-ray AGN at $0.5<z<4.5$: host galaxies dictate dark matter halo mass}
\author[Charutha Krishnan]{Charutha Krishnan$^{1}$\thanks{E-mail: charutha.krishnan@nottingham.ac.uk}, Omar Almaini$^{1}$, Nina A. Hatch$^{1}$, Aaron Wilkinson$^{2}$,
\newauthor David T. Maltby$^{1}$, Christopher J. Conselice$^{1}$, Dale Kocevski$^{3}$, Hyewon Suh$^{4}$,
\newauthor Vivienne Wild$^{5}$\\
$^{1}$School of Physics and Astronomy, University of Nottingham, University Park, Nottingham, NG7 2RD, United Kingdom \\
$^{2}$Department of Physics and Astronomy, Universiteit Gent, Krijgslaan 281, S9, 9000 Gent, Belgium \\ 
$^{3}$Department of Physics and Astronomy, Colby College, Waterville, ME 04901, USA \\ 
$^{4}$Subaru Telescope, National Astronomical Observatory of Japan (NAOJ), National Institutes of Natural Sciences (NINS), \\
650 North A`\={o}h\={o}ku place, Hilo, HI 96720, USA \\ 
$^{5}$School of Physics and Astronomy, University of St Andrews, North Haugh, St Andrews, KY16 9SS, United Kingdom \\
}

\begin{document}
\label{firstpage}
\pagerange{\pageref{firstpage}--\pageref{lastpage}}
\maketitle
\begin{abstract}

We present evidence that AGN do not reside in ``special'' environments, but instead show large-scale clustering  determined by the properties of their host galaxies. Our study is based on an angular cross-correlation analysis applied to X-ray selected AGN in the COSMOS and UDS fields, spanning redshifts from $z\sim4.5$ to $z\sim0.5$. Consistent with previous studies, we find that AGN at all epochs are on average hosted by galaxies in dark matter halos of $10^{12}-10^{13}$ M$_{\odot}$, intermediate between star-forming and passive galaxies. We find, however, that the same clustering signal can be produced by inactive (i.e. non-AGN) galaxies closely matched to the AGN in spectral class, stellar mass and redshift. We therefore argue that the inferred bias for AGN lies in between the star-forming and passive galaxy populations because AGN host galaxies are comprised of a mixture of the two populations. Although AGN hosted by higher mass galaxies are more clustered than lower mass galaxies, this stellar mass dependence disappears when passive host galaxies are removed. The strength of clustering is also largely independent of AGN X-ray luminosity.  We conclude that the most important property that determines the clustering in a given AGN population is the fraction of passive host galaxies. We also infer that AGN luminosity is likely not driven by environmental triggering, and further hypothesise that AGN may be a stochastic phenomenon without a strong dependence on environment.
\end{abstract}


\section{Introduction}
\label{sec:intro}

The astronomical community is in consensus that essentially all massive galaxies host super-massive black holes (SMBHs), that are observed as Active Galactic Nuclei (AGN) during their phases of intense mass accretion. Numerous lines of observational evidence point towards a tight correlation between the evolution of galaxies and that of their SMBHs. For instance, the total star formation rate density and the total AGN accretion density appear to follow similar trends from $z\sim2$ to $z\sim0$ \citep[e.g.,][for a recent review]{Boyle_1998, Franceschini_1999, Silverman_2008, Kormendy_Ho_2013}, implying a link between the mass growth of SMBHs and their host galaxies.  The properties of galaxies are also correlated with their environment \citep[e.g.,][]{Gisler_1978,Dressler_1999, DeLucia_2006, Conselice_2014}, which suggests that supermassive black holes may also be linked to their large-scale
environment. Indeed, several studies have found correlations between the abundance of AGN and their environment. For example, \citet{Kauffmann_2004} showed there is an enhancement of AGN in low-density regions compared to high-density regions in the local Universe, whilst at higher redshift the opposite is found, with an enhanced fraction of AGN in high-redshift protoclusters relative to the field \citep{Lehmer_2009,Digby-North_2010,Krishnan_2017}.

A useful tool to examine the possible connection between AGN and the large scale environment in a statistical manner is the angular and spatial clustering of AGN. As SMBHs populate collapsed dark matter halos in the $\Lambda$CDM paradigm, they can be assumed to reflect the peaks in the spatial distribution of dark matter in the Universe. The 2-point correlation function (2pcf) is the most commonly used tool for large-scale clustering analysis \citep{Peebles_1980}. The 2pcf of galaxies expresses the excess probability of finding pairs of galaxies, above a random distribution. Comparison of the observed 2pcf to that of dark matter from the outputs of detailed dark matter simulations allows the determination of the dark matter halo masses of galaxies hosting AGN. In theory, the typical large-scale environments of AGN can then be inferred as a function of cosmic time, providing potential insights into the connection between AGN and their large-scale environments. 

The 2pcf of quasars has been studied extensively in the literature and are found to reside in $10^{12-12.7}$ M$_{\odot}$ halos out to $z\sim4$ \citep[]{Croom_2005,Shen_2007,Ross_2009,Eftekharzadeh_2015,Ikeda_2015,Garcia-Vergara_2017}. Similarly, lower-luminosity broad line AGN out to $z\sim0.5$ are hosted by dark matter halos of $10^{13}$ M$_{\odot}$ \citep{Miyaji_2011,Krumpe_2012}, and X-ray AGN across a wide range of redshifts ($z\sim0.05$, $z\sim0.1$, $z\sim0.98$, $z\sim1.25$, $z\sim3.4$) inhabit halos of $\sim10^{13}$ M$_{\odot}$ \citep[][respectively]{Powell_2018,Mountrichas_Georgakakis_2012,Koutoulidis_2013,Bradshaw_2011,Allevato_2016}. Many of these studies interpret the $10^{12-13}$ M$_{\odot}$ halo mass as evidence that groups are the ideal environment in which AGN are triggered through galaxy-galaxy interactions \citep[e.g.,][]{Miyaji_2011,Ikeda_2015}. Inconsistent with this interpretation, however, is the lack of evidence for an enhancement of AGN in groups relative to the field \citep{Shen_2007, Arnold_2009, Oh_2014, Tzanavaris_2014}. 

Interpretation of the halo mass derived from clustering is complicated because it may not necessarily represent the environments of AGN, unless AGN are a consistent population inhabiting a single environment. If AGN are comprised of different populations of galaxies, each inhabiting a different environment, then the clustering of AGN will indicate the average of these environments. In fact, recent studies at $z\lesssim1$ suggest that host galaxies play a major role in clustering measurements of AGN. For example, at $z\sim0.7$, \citet{Mendez_2016} find significant differences in the clustering measurements of X-ray, mid-IR and radio AGN, and that these differences are driven by differences in host galaxy properties (such as stellar mass and star-formation rate). Furthermore, at $z<0.1$, \citet{Powell_2018} find that, when accounting for host galaxy properties, AGN occupy dark matter halos consistent with the overall inactive galaxy population. 

Drawing a consistent picture of the clustering of AGN is difficult because previous studies select AGN in various methods with different surveys and telescope sensitivities, thus sampling different distributions of host galaxy properties. Furthermore, the clustering is quantified using differing methods (e.g. angular, projected, real-space correlation functions) and a diverse range of models to estimate halo masses. While several works derive bias values of AGN/quasars using a broad redshift range, it has not been possible as yet for a single study to measure the \emph{evolution} of the AGN bias with redshift from $z\sim4.5$ to $z\sim0.5$, due to the large samples required to obtain reliable bias measurements from the AGN auto-correlation function. In this work, we are able to use the angular cross-correlation function using the more numerous underlying galaxy sample from UDS and COSMOS, as these surveys provide the unique combination of depth and area required to detect large numbers of galaxies out to $z\sim4$. This cross-correlation technique allows us to then reliably infer the auto-correlation functions of AGN, affording us the opportunity to split our X-ray AGN sample into several redshift intervals. In this study we are also able to investigate potential interpretations of the clustering signal of AGN by considering the influence of host galaxy properties such as mass and star-formation characteristics. 

This paper is structured as follows. We describe the data sets used in Section~\ref{sec:data_sample} as well as our sample selection. Section~\ref{sec:clustering} explains the methodology of our clustering analysis, and we present our results in Section~\ref{sec:results}. We adopt a WMAP9 cosmology \citep{Hinshaw_2013}, with $\Omega_\text{m} = 0.3$, $\Omega_\Lambda = 0.7$, and $h = 0.7$. All magnitudes are in the AB system. All X-ray luminosities quoted are calculated in rest-frame bands using a power-law model with a photon index $\Gamma=2$. We note that the effect of Galactic absorption on our fluxes is negligible.

\section{Data sets and sample selection}
\label{sec:data_sample}
In this section we review the data used in our study and describe our sample selection.

\subsection{Description of the data}
\label{sec:data}

We make use of two deep and wide near-infrared surveys for our galaxy samples (Section~\ref{sec:UDS} and Section~\ref{sec:COSMOS}) with Chandra and XMM X-ray coverage.

\subsubsection{UDS DR11 catalogue}
\label{sec:UDS}

The first data set used to form our $K$-band selected galaxy sample is the UKIRT Infrared Deep Sky Survey \citep[UKIDSS,][]{Lawrence_2007} Ultra Deep Survey (UDS). The UDS is a deep photometric survey centred on RA = 02:17:48, DEC = -05:05:57, covering a survey area of $0.62$deg$^2$ after removing masked regions such as bright stars and image artefacts. In this work we use the 11th data release of the UDS (UDS DR11). The  $5\sigma$ limiting depths in $2''$ diameter apertures are $25.6$, $25.1$ and $25.3$ mag in the $J$, $H$, and $K$-bands, respectively \citep[][Almaini et al. in prep]{Hartley_2013}. This catalogue uses deep $Y$-band observations  provided by the VISTA VIDEO survey \citep{Jarvis_2013}, optical observations ($U$, $B$, $V$, $R$, $i'$, $z'$) from the Subaru XMM-Newton Deep Survey \citep[SXDS,][]{Furusawa_2008}, as well as IRAC 3.6$\mu$m and 4.5$\mu$m coverage from the \spitzer\ UDS Legacy Program (SpUDS, PI:Dunlop) and the \spitzer\ Extended Deep Survey \citep[SEDS,][]{Ashby_2013}.

As described in \citet{Simpson_2013}, photometric redshifts were determined by fitting 12-band photometry using a library of templates built from the \citet{Bruzual_Charlot_2003} models with range of ages and metallicities, using a Small Magellanic Cloud extinction law. These photometric redshifts have a normalised median absolute deviation of $\sigma_\text{NMAD} = 0.019$ as compared to $\sim7000$ secure spectroscopic redshifts, with an outlier fraction of 4.5\% (objects with $|z_p - z_s|/(1 + z_s) > 0.15$). Spectroscopic redshifts are used when available. We refer the reader to Almaini et al. (in prep) for further details on the UDS DR11 catalogue.

\subsubsection{COSMOS2015 catalogue}
\label{sec:COSMOS}

The second data set we use is the Cosmic Evolution Survey \citep[COSMOS,][]{Scoville_2007}. COSMOS is a comparable survey to the UDS that reaches shallower depths but covers a larger area of 1.5 deg$^2$ in the UltraVISTA-DR2 region centred on RA = 10:00:28, DEC = +02:12:21. We draw our galaxy sample from \emph{COSMOS2015} published in \citet{Laigle_2016}, with PSF-matched photometry from Subaru SuprimeCam in the $u+$, $B$, $V$, $r$, $i$, and $z++$  bands. UltraVISTA \citep{McCracken_2012} provide NIR photometry in the $Y$, $J$, $H$, and $K_s$ bands. Although UltraVISTA comprises of ``ultra-deep'' stripes over roughly half the area, we limit our sample with $K$-band completeness limits corresponding to shallower regions to maximise number statistics while selecting galaxies uniformly across the field. Subaru Hyper Suprime-Cam (HSC) Y band imaging and \spitzer\ IRAC observations (3.6$\mu$m and 4.5$\mu$m bands) are also included. 

We use photometric redshifts from \citet{Laigle_2016} that have made use of \citet{Bruzual_Charlot_2003} models, as well as templates of spiral and elliptical galaxies from \citet{Polletta_2007}. As described in \citet{Laigle_2016}, the uncertainty of the photometric redshifts varies from $\sigma_\text{NMAD} = 0.007$ with an outlier fraction of 0.5\% (for bright, low redshift, star-forming galaxies) to $\sigma_\text{NMAD} = 0.021$  with an outlier fraction of 13.2\% (for the $z>3$ sample), where outliers are objects with $|z_p - z_s|/(1 + z_s) > 0.15$. We use spectroscopic redshifts when available. The reader is referred to \citet{Laigle_2016} for further details on \emph{COSMOS2015}. 

\subsubsection{X-ray AGN counterpart catalogues}
\label{sec:X-ray}

For both UDS and COSMOS, we use AGN catalogues that identify optical/NIR counterparts for X-ray sources using a maximum likelihood approach. The UDS field has wide but shallow XMM observations, as well as deep Chandra coverage of a smaller fraction of the field. Therefore we use the Chandra counterparts, but supplement this with XMM counterparts outside the Chandra covered region.

The X-UDS observations were carried out over $1.25$ Ms with Chandra's Advanced  CCD  Imaging  Spectrometer \citep[ACIS,][]{Garmire_2003}, 25 ACIS-I pointing positions that cover a total area of roughly $35' \times 25'$ in size. The final X-ray source catalogue, presented in \citet{Kocevski_2018}, amounts to $868$ sources. Counterparts to these X-ray sources were matched to the CANDELS $H$-band and UDS DR10 $K$-band catalogues using the likelihood ratio technique of \citet{Sutherland_Saunders_1992}, following the method outlined in \citet{Civano_2012}. Spectroscopic redshifts are available for $\sim400$ sources. See Hasinger et al. (in prep) for further details on the counterpart matching procedure. 

The UKIDSS UDS survey is also observed by the Subaru-XMM-Newton Deep Survey (SXDS), mapped by seven pointings with XMM-Newton covering $0.2-10$ keV. The X-ray source catalogue is presented in \citet{Ueda_2008}, amounting to 1245 sources. XMM counterparts have been obtained using the likelihood ratio method to $R$-band, $3.6\mu$m, near-UV, and $24\mu$m source catalogues. Spectroscopic observations allow the identification of 597 out of 896 total AGN. The remaining AGN have redshifts derived using 15 band photometry, where separate SED templates of QSOs and galaxies are applied to each counterpart. See \citet{Akiyama_2015} for further details.

The Chandra COSMOS-Legacy is the product of 4.6 Ms of Chandra observations over the 2.2 deg$^2$ COSMOS area. The X-ray source catalogue is described in \citet{Civano_2016} and amounts to 4016 sources. \citet{Marchesi_2016} identify counterparts to the X-ray sources using the approach of \citet{Civano_2012}, making use of three different bands: i, Ks and 3.6 $\mu$m. Of the 4016 sources, $\sim97\%$ have optical/near-infrared counterparts, and $\sim54\%$ have spectroscopic redshifts. See \citet{Marchesi_2016} for further details on the counterpart catalogue.

These surveys are comparable in the hard band ($2-10$ keV) and we thus only study AGN detected in this band. Photometric and spectroscopic redshifts of the X-ray sources in X-UDS, SXDS and Chandra COSMOS-Legacy were provided by the most likely counterpart. In all cases we use the spectroscopic redshift when available. 

\subsubsection{Supercolour catalogues}
\label{sec:SC}

We determine the spectral class of our AGN and galaxy samples using supercolour classifications that utilise a Principal Component Analysis (PCA) of the optical/near-infrared photometric data in the UDS and COSMOS fields  \citep{Wild_2014, Wild_2016}.  The eigenvectors for the PCA are built from a library of \citet{Bruzual_Charlot_2003} spectral synthesis models with a variety of ages, metallicities, star formation histories and dust contents. The top three eigenvectors are found to represent $> 99.9\%$ of the variance in the model SED library and the shape of an observed galaxy SED can be represented by the weights of these eigenvectors, termed ``super-colours'' (SC1, SC2 and SC3). This method separates a tight red-sequence from star-forming galaxies and also identifies rarer populations such as post-starburst (PSB) and dusty star-forming galaxies, depending on where they lie in super-colour space. We use the updated boundaries presented in Wilkinson et al. (in prep) to classify our galaxies and AGN hosts as either passive (red + PSB) or star-forming (SF + dusty). The spectral classes and their boundaries, for both the UDS and COSMOS fields, are shown in Table~\ref{tab:SC_boundaries}.

\begin{table*}
\begin{tabular}{|l|l|l||}
\hline
Spectral class 	& UDS 					& COSMOS					\\
\hline	
SF		& $SC2 < 0.4   \times SC1+10.86$	& $SC2 < 0.38  \times SC1+11.67$		\\ 
		& $\cap \: SC2 < 0.783 \times SC1+14.83$& $\cap \: SC2 < 0.82  \times SC1+16.39$	\\ 
		& $\cap \: SC2 > -0.2  \times SC1-13.75$& $\cap \: SC2 > -0.18 \times SC1-12.12$	\\ 
Red		& $SC2 > 0.783 \times SC1+14.83$	& $SC2 > 0.82  \times SC1+16.39$		\\ 
		& $\cap \: SC2 < -0.34 \times SC1+3.19$ & $\cap \: SC2 < -0.26 \times SC1+3.94 $	\\ 
PSB		& $SC2 > 0.4   \times SC1+10.86$	& $SC2 > 0.38  \times SC1+11.67$		\\ 
		& $\cap \: SC2 > -0.34 \times SC1+3.19$ & $\cap \: SC2 > -0.26 \times SC1+3.94 $	\\ 
Dusty		& $SC2 < -0.2  \times SC1-13.75$	& $SC2 < -0.18 \times SC1-12.12$		\\ 
		& $SC1<-20$				& $SC1<-19.5$                                                            \\
\hline
\end{tabular}
\caption{\label{tab:SC_boundaries} Spectral class definitions according to supercolour boundaries.}
\end{table*}

As the broadband filters are slightly different between the UDS and COSMOS fields, the PCA analysis is performed separately on each field resulting in different eigenvectors. Wilkinson et al. (in prep) use models and spectroscopically classified galaxies to ensure consistency in the classification boundaries in SC space between the two fields. Due to deeper data in the UDS field, galaxies can robustly be classified to $z=3$, but galaxies in COSMOS are limited to $z=2.5$. Therefore we adopt a uniform cut in redshift at $z<2.5$ when investigating the role of host galaxy properties. 

The stellar masses used in this work are determined using this supercolour technique, assuming a \citet{Chabrier_2003} initial mass function (IMF) and \citet{Bruzual_Charlot_2003} stellar population templates (see \citealt{Wild_2016} and Wilkinson et al. in prep for more details). The uncertainty from the supercolour Bayesian stellar mass fitting is typically $0.1$ dex, with no indication of additional uncertainty for AGN. This uncertainty allows for the degeneracy between fitted parameters and the photometric redshift uncertainties. Further discussion of the supercolour stellar mass uncertainties can be found in \citet{Almaini_2017}. Based on this work, we expect our results to be robust to known sources of random and systematic errors.

\subsection{Sample selection}
\label{sec:sample}

\subsubsection{Galaxy samples}
\label{sec:galaxy_sample}

\begin{figure}
\includegraphics[width=\columnwidth,trim=0.5cm 0.0cm 1.2cm 0cm]{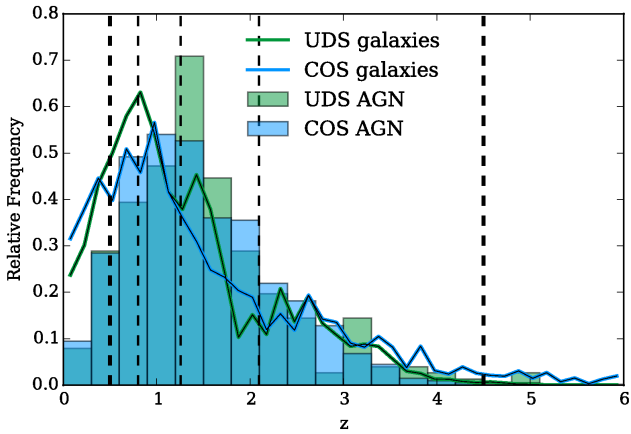}
\vspace*{-0.5cm}
\caption{Redshift distribution of the AGN in the UDS and COSMOS fields are shown by the green and blue histograms, respectively. The galaxy redshift distributions within these two fields are shown by the solid lines. The black dashed lines represent our redshift intervals of $0.5<z<0.8$, $0.8<z<1.3$, $1.3<z<2.1$, $2.1<z<4.5$, and the thicker lines denote the redshift limits of our study.}
\label{fig:z_dist}
\end{figure}

To study the evolution of clustering of AGN as a function of cosmic time, we split our AGN and galaxy samples in redshift intervals of $0.5<z<0.8$, $0.8<z<1.3$, $1.3<z<2.1$, $2.1<z<4.5$ corresponding to equal cosmic time intervals of $1.8$ Gyr. In order to maximise the quality of our galaxy sample, we apply a maximum limit to the minimum $\chi^2$ values associated with fitting photometric redshifts. We also apply a $K$-band magnitude limit of $23.7$ to our galaxy sample. After applying these quality cuts, our galaxy sample consists of $\sim60000$ and $\sim170000$ galaxies between $0.5<z<4.5$ in the UDS and COSMOS fields, respectively. In addition, we use the methodology from \citet{Pozzetti_2010} to apply redshift-dependent $90\%$ mass completeness limits to the galaxy sample. We do not apply a mass completeness cut to our AGN sample to maximise the sample size, but we note that $95\%$ of our AGN are above the $90\%$ mass completeness limit of $10^{10.2}$\,M$_{\odot}$ at $z=2.5$.  The clustering measurements are robust to more conservative cuts in the $K$-band magnitude and mass completeness, as results are consistent within error-bars, albeit with larger uncertainties. The redshift distribution of the galaxy samples are shown by the solid lines in Figure~\ref{fig:z_dist}.

\subsubsection{AGN samples}
\label{sec:AGN_sample}

It is important to select AGN detected to a uniform flux limit to ensure that the clustering measurements are not biased by the varying source density with exposure. We select our hard band AGN to a flux limit of $3.55\times10^{-15}$ erg s$^{-1}$ cm$^{-2}$. AGN outside the area corresponding to this flux limit are removed, as well as AGN within the region that have fainter fluxes than the flux limit. This limit was chosen to maximise the number of AGN, and applied to AGN in both the UDS and COSMOS fields. In the UDS, the X-UDS coverage is deeper but limited to the central $\sim0.33$ deg$^2$. Outside this Chandra-covered region, we therefore supplement our X-UDS data with XMM-Newton data.

\begin{figure*}
    \centering
    \begin{minipage}{.48\textwidth}
        \centering
        \includegraphics[width=\columnwidth,trim={0.25cm 0 0.9cm 0},clip]{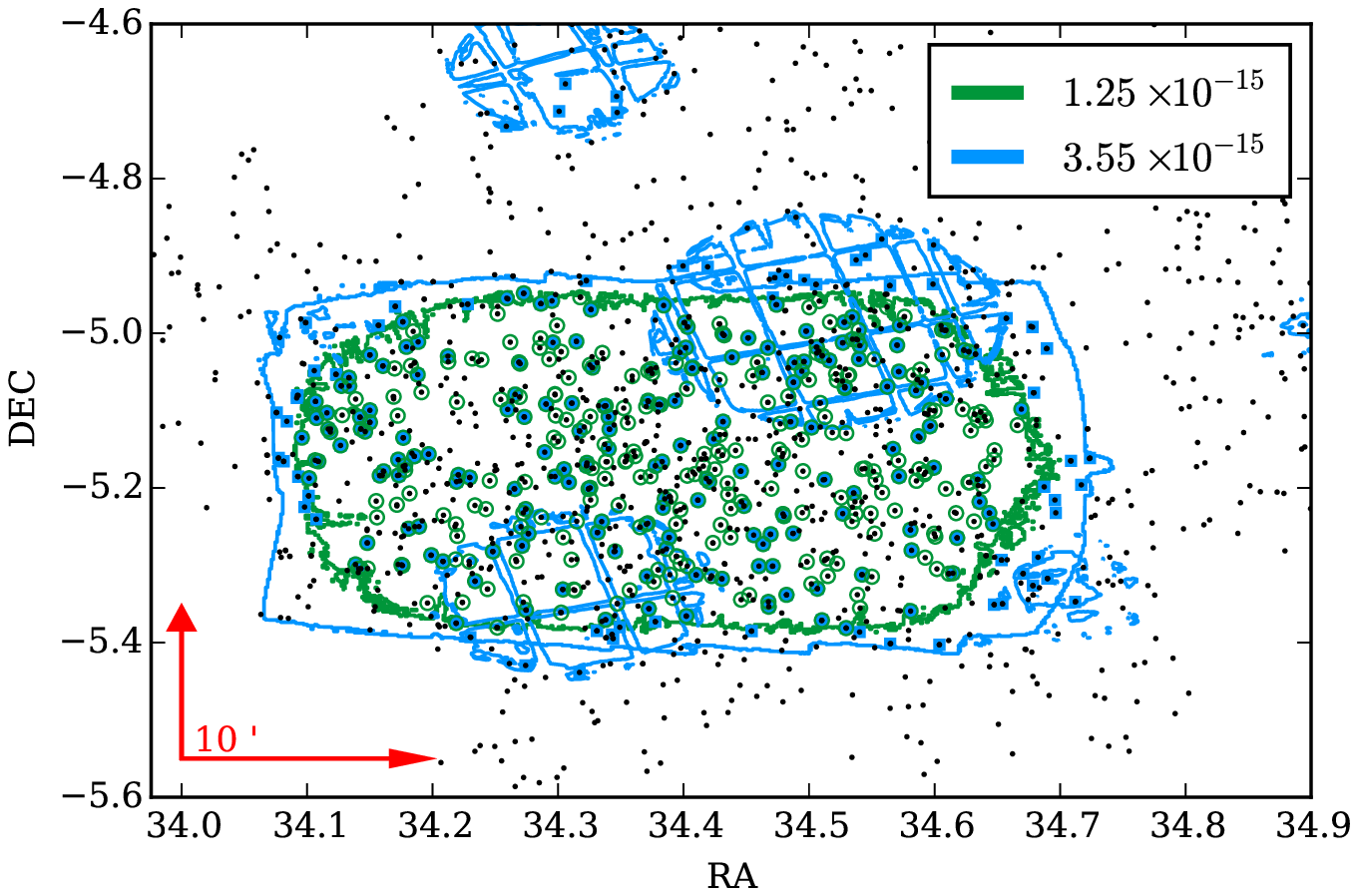}       
    \end{minipage}
    \begin{minipage}{.48\textwidth}
        \centering
        \includegraphics[width=\columnwidth,trim={0.25cm 0 0.9cm 0},clip]{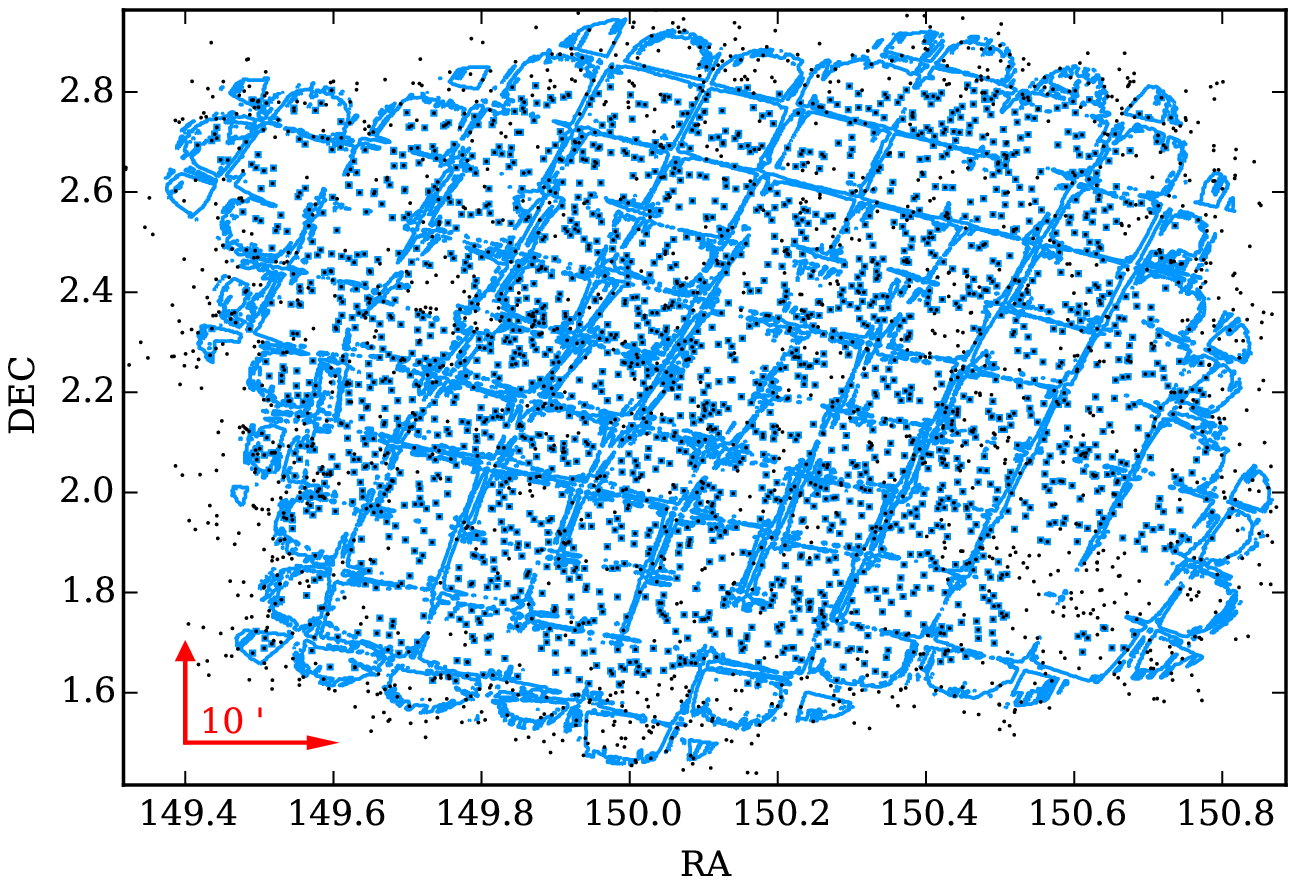}        
\end{minipage}
\vspace*{-0.3cm}
\caption{Hard-band X-ray AGN in UDS Chandra and XMM (left) and COSMOS (right). All X-ray sources with optical/near-infrared counterparts are shown as black points, AGN with a flux limit of $\geq1.25\times10^{-15}$ erg s$^{-1}$ cm$^{-2}$ and $\geq 3.55\times10^{-15}$ erg s$^{-1}$ cm$^{-2}$ are highlighted in green circles and blue squares, respectively. The contours corresponding to these flux limits are also shown in the respective colours. Note that the larger size of the COSMOS field may give an appearance of a higher density of sources, but this does not reflect reality. For reference, the red arrows in the bottom left corner of the two plots show a scale of 10'.}
\label{fig:UDS_COS_flux_lim}
\end{figure*}

While the UDS achieves greater depths in both $K$-band and X-rays, the larger number of AGN in the COSMOS field effectively dictates our flux limit. To ensure a consistent depth, we therefore discard the fainter data in the UDS and adopt a shallower X-ray flux limit and $K$-band magnitude limit. We also make maximum use of the UDS field by computing the optimum flux limit for the UDS field independently ($1.25\times10^{-15}$ erg s$^{-1}$ cm$^{-2}$) and measuring the clustering of AGN in this field. The results are consistent with the measurements using a flux limit of $3.55\times10^{-15}$ erg s$^{-1}$ cm$^{-2}$, with smaller uncertainties. The redshift distributions of the AGN selected to the latter flux limit are shown in Figure~\ref{fig:z_dist}. Figure~\ref{fig:UDS_COS_flux_lim} shows the spatial distribution of the AGN in UDS (left) and COSMOS (right). The flux limit contours optimised for UDS and COSMOS are overlaid in green and light blue respectively. AGN selected to these flux limits are highlighted by the respective colours. 

\begin{figure}
\includegraphics[width=\columnwidth,trim={0.25cm 0 0.9cm 0},clip]{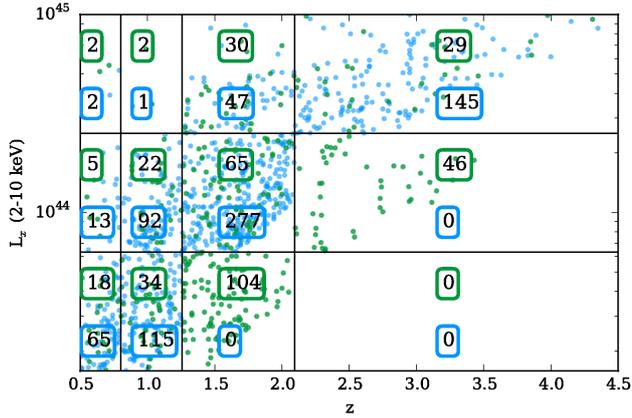}       
\vspace*{-0.3cm}
\caption{The number of AGN in each luminosity and redshift bin in the UDS (green boxes) and COSMOS (blue boxes) fields. We apply a flux limit computed with the lower end of the luminosity bin and the median redshift of the AGN within the bin. The diagonal line traced out by missing points in the bottom right hand corners of the boxes correspond to the sources that have been removed as their fluxes were lower than the applied flux limit. }
\label{fig:Lx_vs_z}
\end{figure}

In our study of the links between clustering and AGN luminosity, we divide AGN into X-ray luminosity and redshift bins (see boxes in Figure~\ref{fig:Lx_vs_z}). To  the AGN in each bin, we apply a flux limit computed with the lower end of the luminosity bin and median redshift of the AGN within the bin. The diagonal lines traced out by missing points in the bottom right hand corners of the boxes in Figure~\ref{fig:Lx_vs_z} correspond to the sources that have been removed as their fluxes were lower than the applied flux limit. 

\section{Clustering methods}
\label{sec:clustering}

In this section, we first measure the angular two point correlation function of AGN (Section~\ref{sec:AGN_clustering}) and that of the underlying dark matter (Section~\ref{sec:DM_clustering}). We then measure the strength of clustering, using the ``bias'' parameter we describe in Section~\ref{sec:bias_fitting}, by scaling the dark matter CF to the AGN CF and minimising $\chi^2$. In this section we also derive the dark matter halo masses of AGN as a function of redshift. 

\subsection{Two point correlation functions}
\label{sec:correlation_fns}

\subsubsection{AGN correlation functions}
\label{sec:AGN_clustering}

The most commonly used statistical estimator of galaxy clustering uses the two-point auto-correlation function (ACF). The angular ACF, $w(\theta)$, measures the excess probability, above a random distribution, of finding a galaxy at an angular separation $\theta$ from another galaxy. We use the \citet{Landy_Szalay_1993} estimator, described by 
\begin{equation}
w(\theta)=\frac{DD(\theta)-2DR(\theta)+RR(\theta)}{RR(\theta)},
\label{eq:auto}
\end{equation}
\noindent
where $DD(\theta)$, $DR(\theta)$ and $RR(\theta)$ are the galaxy-galaxy, galaxy-random and random-random normalised pair counts, respectively. We choose this estimator because it is more robust to effects that can affect clustering measurements, such as the size of the random catalogue and edge corrections \citep{Coil_2013}.

While the clustering of AGN and quasars have been studied using the ACF, this method requires large sample sizes to provide tight constraints on AGN host halo masses. As our AGN sample sizes are limited, we can make use of a close cousin of the ACF; the two-point cross-correlation function (CCF). We measure the CCF of the AGN with respect to $K$-band selected galaxy samples provided by the UDS and COSMOS surveys. 

We cross-correlate our AGN ($D_{\rm{A}}$) with a full volume-limited 90\% mass-complete tracer galaxy population ($D_{\rm{G}}$), using:
\begin{equation}
w(\theta)=\frac{D_{\rm{A}}D_{\rm{G}}(\theta)-D_{\rm{A}}R_{\rm{G}}(\theta)-D_{\rm{G}}R_{\rm{A}}(\theta)+R_{\rm{A}}R_{\rm{G}}(\theta)}{R_{\rm{A}}R_{\rm{G}}(\theta)},
\label{eq:cross}
\vspace{0.15cm}
\end{equation}
\noindent
where each term is normalised by the total pair counts. $R_{\rm{A}}$ and $R_{\rm{G}}$ denote the random source catalogues that populate the regions from which we select AGN (see Figure~\ref{fig:UDS_COS_flux_lim}) and tracer galaxies respectively. For the tracer galaxy catalogues, we populate random catalogues corresponding to the ``good'' regions of UDS and COSMOS with artefacts such as stars and cross-talk masked out. For our AGN catalogues in UDS and COSMOS, we mask for the bad regions of the relevant optical/near-infrared images, as well as the regions of the X-ray image for which the flux limit is shallower than $3.55\times10^{-15}$ erg s$^{-1}$ cm$^{-2}$. For the study of luminosity-dependent clustering, we compute different random catalogues for each of the flux limits derived from the different luminosity and redshift bins. These random catalogues map out the same regions as each of our galaxy catalogues, and the total number of randoms is $50000$ and $100000$ in the UDS and COSMOS fields respectively. The ratio of randoms to data is roughly consistent for each redshift bin (of order 300:1).

In order to ensure that our clustering measurements are reliable, we impose a lower limit of 30 AGN in a given sample to qualify for our analysis. 

As the clustering measured using the CCF is underestimated due to the limited observed field size, we apply a correction factor for the integral constraint $C$ following \citet{Roche_Eales_1999}, 
\begin{equation}
C_{\rm{CCF}}=\frac{\sum R_{\rm{A}}R_{\rm{G}}(\theta) w(\theta)}{\sum R_{\rm{A}}R_{\rm{G}}(\theta)}.
\label{eq:IC_CCF}
\end{equation}

The integral constraint for the ACF of our tracer sample is similarly defined as,
\begin{equation}
C_{\rm{ACF}}=\frac{\sum R_{\rm{G}}R_{\rm{G}}(\theta) w(\theta)}{\sum R_{\rm{G}}R_{\rm{G}}(\theta)}.
\label{eq:IC_ACF}
\end{equation}

\noindent We assume that $w(\theta)$ of AGN traces the angular correlation function of the underlying dark matter distribution, following the method of \citet{Hartley_2013} and \citet{Wilkinson_2017}.

\subsubsection{Dark matter correlation functions}
\label{sec:DM_clustering}

In order to interpret our angular correlation functions, we compute the angular correlation function of the matter distribution (dominated by dark matter) over the same volume as our tracer sample. We do this following the formalism of \citet{Smith_2003} to compute the non-linear dark matter power spectrum at the mean redshift of the sample. This is then Fourier transformed to obtain the 3D non-linear dark matter angular correlation function; i.e. the sum of the correlation functions in the 1-halo regime (pair counts within the same halo) and the 2-halo regime (pair counts in different halos). Finally, we project this onto 2D using the redshift distribution of the sample and the relativistic Limber equation \citep{Limber_1954}, following \citep{Peebles_1980, Magliocchetti_Maddox_1999}. We refer the reader to Section~4 of \citet{Magliocchetti_Maddox_1999} for a clear account of this implementation. To determine the redshift distribution, we use a top hat redshift probability function for both our AGN and tracer samples, since the AGN redshift probability distributions may not be as reliable as those of the tracer sample. The uncertainty on these redshift distributions is the main limitation in projecting using the Limber equation. In addition, the slope of the real-space correlation function must be known, and a cosmology must be assumed. Although the projection depends on the assumed cosmological parameters, this is a systematic effect such that the comparison of clustering between different populations in the same cosmology is unaffected.

This routine therefore allows us to obtain synthetic w$_\text{dm}(\theta$) from the halo model that can be fit to the observed AGN correlation functions.

\subsection{Bias fitting and halo masses}
\label{sec:bias_fitting}

Galaxies are biased tracers of the underlying dark matter density field. The linear galaxy bias parameter $b$, defined as the ratio of the overdensity of galaxies to the ($\delta_g$) to the mean overdensity of matter, 
\begin{equation}
b = \delta_g/\delta,
\end{equation}
\noindent
where $\delta$ is the overdensity, given by  $\delta = \rho/\bar{\rho} - 1$ where $\bar{\rho}$ is in turn the mean mass density \citep[e.g.,][]{Peebles_1980}. Therefore galaxies with higher bias have a higher degree of clustering and are more likely to be found near the highest density peaks in the dark matter mass distribution.

We therefore use the bias $b$ to indicate how strongly clustered our AGN sample is with respect to the underlying dark matter distribution. On linear scales, we can compute this parameter as the square root of the measured observed galaxy correlation function divided by the 2D dark matter correlation function following the definition, 
\begin{equation}
w_{\rm{obs}}(\theta)=b^2\times w_{\rm{dm}}(\theta),
\label{eq:wobs}
\end{equation}
\noindent
where $w_{\rm{obs}}$ denotes the observed AGN cross-correlation function and $w_{\rm{dm}}$ denotes the projected correlation function of the dark matter distribution \citep[e.g.,][]{Benson_2000}. 

We therefore fit the dark matter correlation functions multiplied by $b^2$ to the observed AGN correlation function, and calculate the optimum value of $b$ using $\chi^2$ minimisation with weights corresponding to the inverse of the uncertainties on the observed correlation function. These uncertainties are calculated using the bootstrap method with 50 repetitions. We therefore resample the tracer/target sample with replacement $50$ times, and evaluate $w(\theta)$ for each of the $50$ bootstrap samples. The uncertainties are then given by the standard deviation of resampled values of $w(\theta)$. To obtain a combined bias measurement using two separate fields (UDS and COSMOS), we obtain individual $\chi^2$ measurements of the two fields (using two independent AGN and dark matter correlation functions) and minimise the total $\chi^2$, taking into account the correlated uncertainties. 
 
\begin{figure*}
\includegraphics[width=\textwidth,trim=1.5cm 1.0cm 2.0cm 2cm]{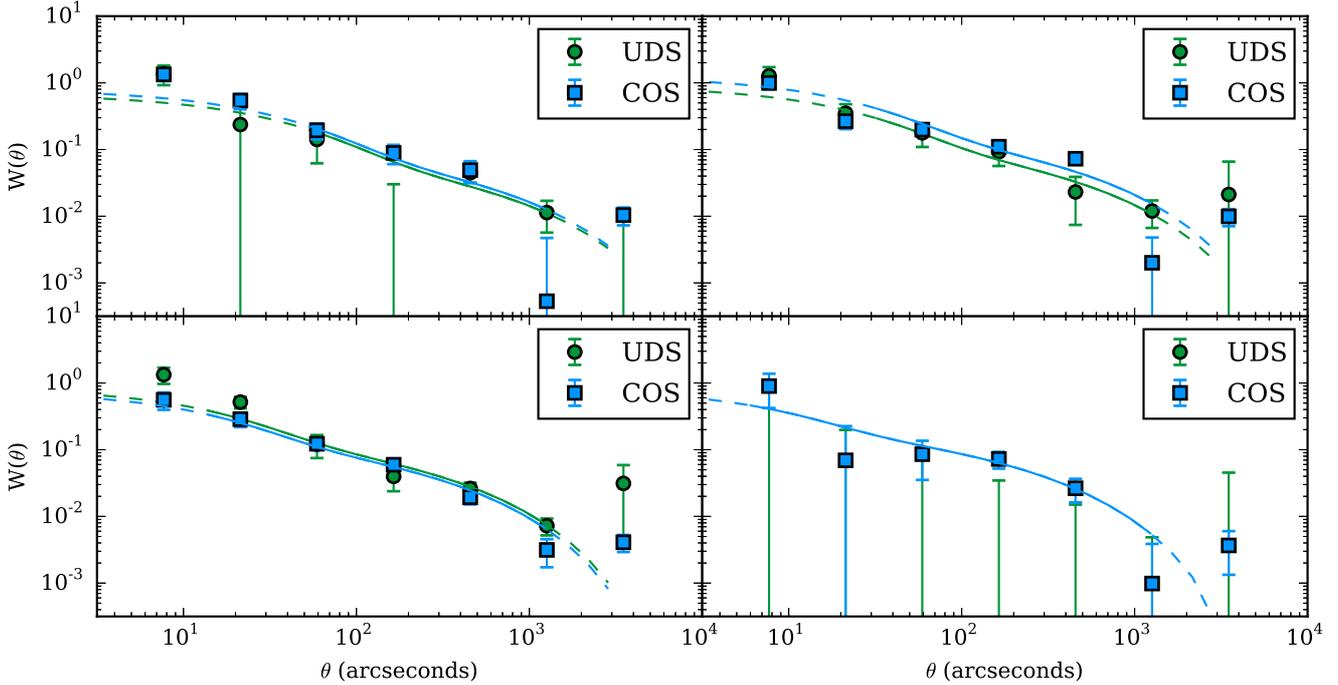}
\vspace*{0cm}
\caption{Inferred auto-correlation functions of X-ray AGN in the UDS (green) and COSMOS (light blue), as a function of redshift. We present the bias measurement with a flux limit optimised separately for each field. Uncertainties are derived from the standard deviation in resampled $w(\theta)$. We note that low number statistics can formally lead to negative $w(\theta)$ in certain bins, resulting in an underestimation of the bias. The combined measurements in these cases will not be affected by this problem as these points will have larger uncertainties and hence lower weights in the $\chi^2$ minimisation. The dashed lines represent the fits of the dark matter correlation functions to the observed correlation functions (scaled by the square of the bias), and the over-plotted solid lines highlight the scales over which the observed correlation function is used in the fitting routine.}
\label{fig:ccf}
\end{figure*}

We assume that both AGN and tracer galaxy populations trace the dark matter distribution, and that both populations are linearly biased. This implies that we expect the correlation function to be well described by linear gravity theory ($w_{\rm{linear}}$). This assumption holds true at large scales where both the linear ($w_{\rm{linear}}$) and non-linear ($w_{\rm{non-linear}}$) correlation functions are mutually consistent, since the non-linear terms are negligible. However, the two CFs deviate at small scales, \citep{Zehavi_2005}, so it is no longer appropriate to assume that the galaxy population traces the dark matter distribution. We therefore adopt a lower limit to $\theta$ corresponding to $w_{\rm{non-linear}} = 3\times w_{\rm{linear}}$ between the linear and non-linear regimes in order to accurately constrain our bias measurements, following \citet{Wilkinson_2017}. The angular scale that $3 \times w_{\rm{linear}}$ corresponds to varies with redshift. In Figure~\ref{fig:ccf}, this scale is shown as the minimum scale for which the solid lines are drawn. We choose this limit as a trade-off between minimising the non-linear effects on small scales as well as maximising the scales over which we can use the correlation functions. We also employ an upper limit and do not consider pair counts at scales larger than $\theta=0.4$ degrees as the finite field of view results in unreliable $w(\theta)$ measurements at large $\theta$.     

We thus determine the absolute bias of the AGN cross-correlated with tracer galaxies (CCF bias, $b_{\rm{AG}}$) and of the tracer galaxy auto correlation function (ACF bias, $b_{\rm{G}}$) with respect to the dark matter. These two absolute bias measurements can then be used to infer the relative bias of the target AGN sample $b_{\rm{A}}$ following $b_{\rm{A}}={b^2_{\rm{AG}}}/{b_{\rm{G}}}$.

The ACF of the target AGN sample can be inferred by multiplying the cross-correlation functions by $(b^2_{\rm{AG}}/b^2_{\rm{G}})$. These inferred ACFs of UDS and COSMOS X-ray AGN are plotted in Figure~\ref{fig:ccf} along with their fitted bias.

Finally, we obtain dark matter halo mass estimates using the formalism of \citet{Mo_White_2002}. The bias of dark matter halos is dependent on its mass and given epoch. We thus assign dark matter halo masses to our AGN samples by matching our bias measurements at a given redshift to the bias of dark matter halos of various masses.

\section{Results}
\label{sec:results}

\subsection{Interpreting the bias of X-ray AGN}
\label{sec:AGN_lse}

\begin{figure}
\includegraphics[width=\columnwidth,trim=0.5cm 0.0cm 1.2cm 0cm]{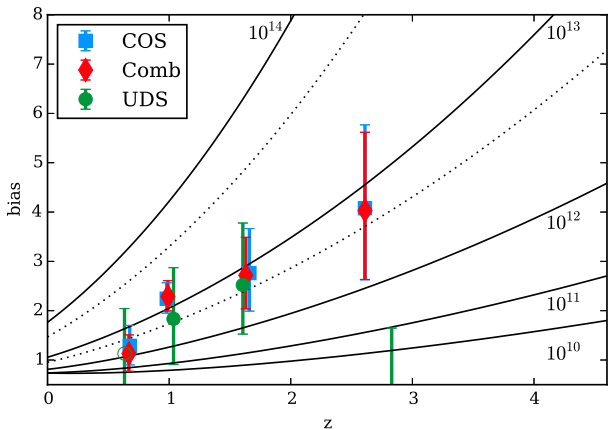}
\vspace*{-0.5cm}
\caption{Redshift evolution of AGN bias in the UDS (green) and COSMOS (light blue). We present the bias measurement with a flux limit chosen to maximise the number of AGN in each field separately, as well as the combined measurement in red optimised for both fields. If fewer than 50 AGN are present in the sample, we plot them in open symbols as these points are potentially less reliable. The redshift evolution of dark matter halos corresponding to given masses are shown by the solid black lines, and are annotated by their corresponding halo masses in solar mass units. The dotted black lines show the evolution of $5\times10^{12}$\,M$_{\odot}$ and $5\times10^{13}$\,M$_{\odot}$. On average, therefore, AGN appear to inhabit $10^{12-13}$\,M$_{\odot}$ halos with no evolution in redshift.}
\label{fig:bias_z}
\end{figure}

We calculate cross-correlation functions in the UDS and COSMOS fields to obtain measurements of the bias and infer the dark matter halo masses of galaxies hosting X-ray AGN as a function of redshift. We plot our measurements in Figure~\ref{fig:bias_z}, where the individual COSMOS and UDS measurements are presented, in addition to the combined estimate of the bias. As shown by Figure~\ref{fig:bias_z}, the clustering of X-ray AGN suggest that they preferentially reside in ``group-like'' environments of $10^{12-13}$\,M$_{\odot}$ irrespective of redshift, in agreement with previous studies \citep{Croom_2005,Ross_2009,Shen_2009, Allevato_2011, Krumpe_2012, Allevato_2016, Powell_2018}.   

\begin{figure}
\includegraphics[width=\columnwidth,trim=0.5cm 0.0cm 1.2cm 0cm]{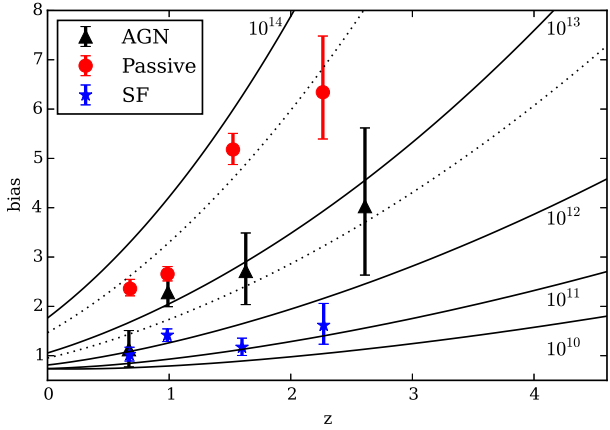}
\vspace*{-0.5cm}
\caption{Redshift evolution of the bias of AGN (black), along with SF (blue) and passive (red) galaxies, as defined by the supercolour technique, matched in mass and redshift. X-ray AGN inhabit intermediate halo masses relative to star-forming and passive galaxies of the same mass distribution. The solid lines denote the evolution of dark matter halos, as in Figure~\ref{fig:bias_z}.}
\label{fig:bias_AGN_SF_pass_z}
\end{figure}

In Figure~\ref{fig:bias_AGN_SF_pass_z}, we compare the bias of AGN to that of star-forming and passive galaxy populations in the UDS and COSMOS fields, selected using the supercolour technique (see Section~\ref{sec:SC} and Wilkinson et al. in prep for more details), matched in mass and redshift distributions to the AGN population\footnote{We note that Wilkinson et al. (in prep) use the full redshift probability distribution to project the 3D dark matter correlation function, whereas we use a top hat redshift probability distribution for consistency with the method we use for our AGN sample.}. As this technique is only reliable out to $z\sim2.5$, we truncate our final redshift bin to $z=2.5$. We find that AGN are more clustered than star-forming galaxies, but less clustered than passive galaxies of the same mass. As it is well known that AGN are a composite population of star-forming and passive galaxies \citep[e.g.,][]{Aird_2017}, the clustering signal may indicate this is a mixed population, or possibly a population transitioning from star-forming to passive.

\begin{figure}
\includegraphics[width=\columnwidth,trim=0.5cm 0.0cm 1.2cm 0cm]{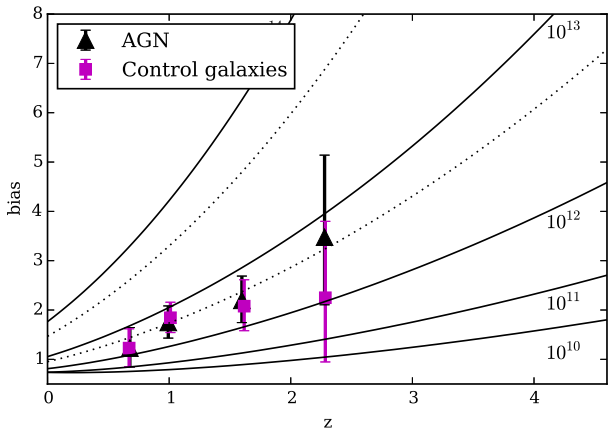}
\vspace*{-0.5cm}
\caption{Redshift evolution of bias of X-ray AGN (for which supercolour classifications are available) in black triangles, compared to a control sample of inactive galaxies (in magenta squares), matched in stellar mass, spectral class (from supercolours), and redshift. The star-formation characteristics (i.e. spectral class) of star-forming galaxies in the non-AGN sample matches that of Figure~\ref{fig:bias_AGN_SF_pass_z}. The solid lines denote the evolution of dark matter halos, as in Figure~\ref{fig:bias_z}. The clustering of AGN is entirely consistent with the clustering of inactive galaxies with similar host galaxy properties.}
\label{fig:AGN_inact_gal}
\end{figure}

To explore this issue, we first classify our AGN host galaxies (for which supercolour classifications are available) into passive and star-forming. Properties of these AGN, such as the median X-ray luminosity and passive host fraction, are summarised in Table~\ref{tab:AGN_details}. We compare the clustering of these AGN to the clustering of an analogous mixed population of star-forming and passive inactive galaxies. We ensure that these inactive galaxies are matched in terms of redshift, spectral class as determined by their supercolours (SC1, SC2 and SC3), and stellar mass. To construct this matched control sample, for each AGN, we identify the closest matching galaxy in a 3-dimensional space of stellar mass, spectral class, and redshift (resulting in a sample of the same size as AGN). The precise matching of control galaxies to AGN results in identical distributions of mass, spectral class, and redshift, all consistent with being drawn from the same underlying distribution ($p=0.99999998$, as probed by a Kolmogorov-Smirnov test). We calculate the bias of this mixed population using the same method applied to the AGN (i.e. by inferring ACF from CCF) and present the results in Figure~\ref{fig:AGN_inact_gal}, showing that the AGN bias measurements in all redshift bins are entirely consistent with inactive galaxies of similar mass and spectral class.

The clustering of the non-AGN control galaxy population is intermediate between passive and star-forming galaxies (see Figure~\ref{fig:bias_AGN_SF_pass_z}), as expected for a mixed sample, with an average clustering signal consistent with ``group'' mass halos. It is therefore possible that the intermediate clustering signal of AGN (corresponding to a halo mass of $10^{12-13}$ M$_{\odot}$) is produced by an averaging of the clustering signal from the mixed host galaxy population, and that the clustering of AGN predominantly reflects the mix of passive and star-forming host galaxies that occupy a range of different environments. 

\subsection{The role of host galaxy properties on AGN clustering}
\label{sec:stellar_mass}

\begin{figure*}
    \centering
    \begin{minipage}{.48\textwidth}
        \centering
        \includegraphics[width=\columnwidth,trim={0.75cm 0 0.9cm 0},clip]{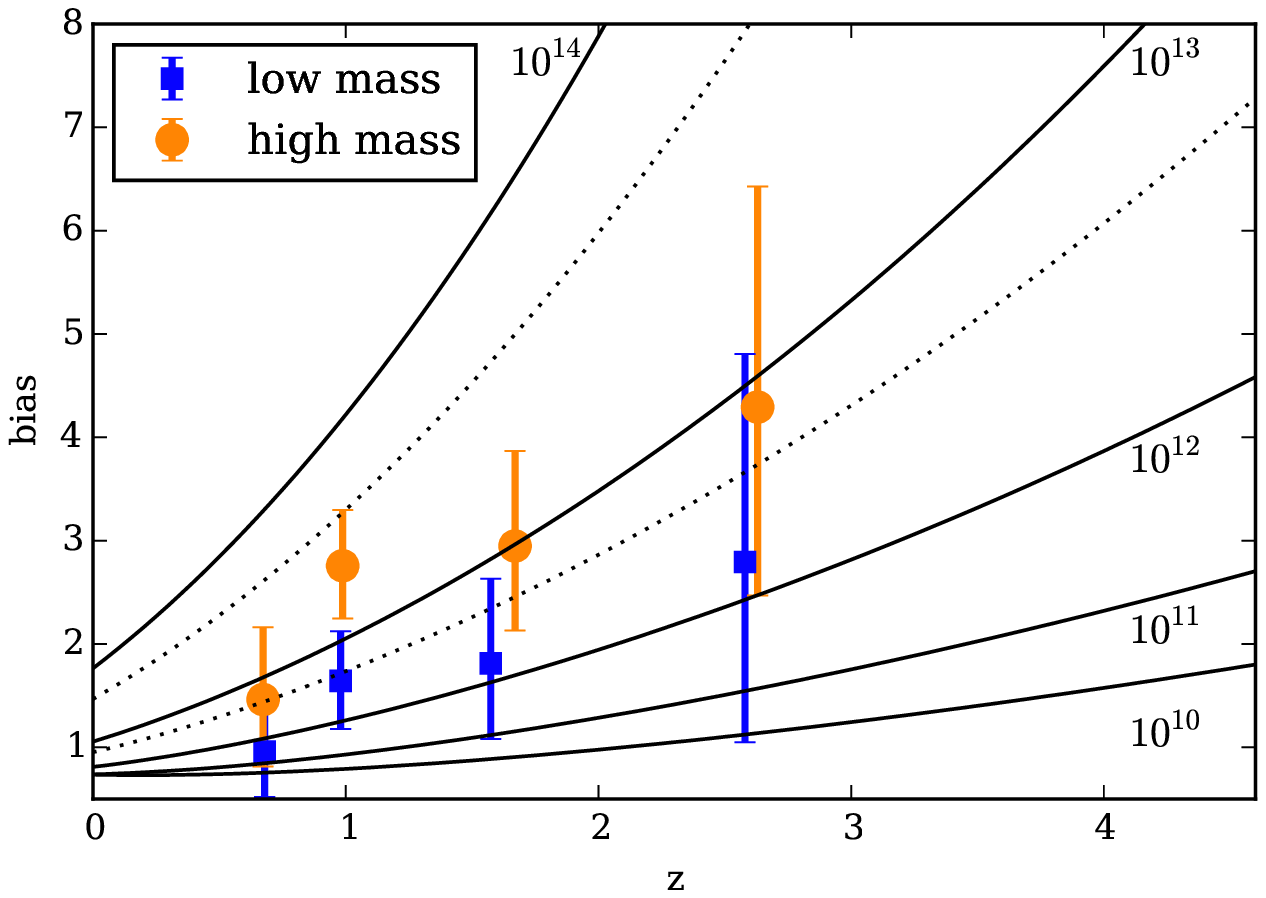}       
    \end{minipage}
    \begin{minipage}{.48\textwidth}
        \centering
        \includegraphics[width=\columnwidth,trim={0.25cm 0 1.4cm 0},clip]{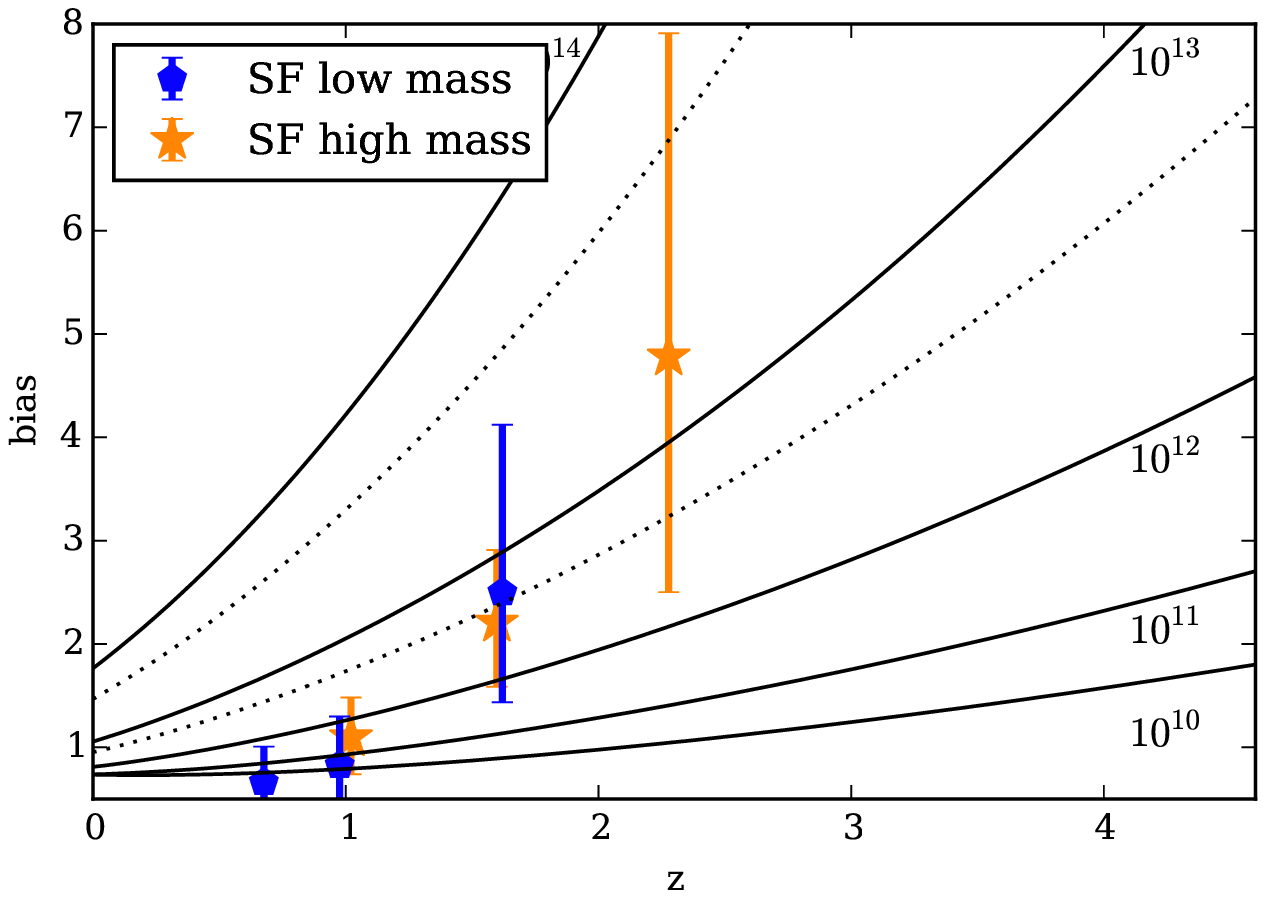}        
\end{minipage}
\vspace*{-0.3cm}
\caption{\textbf{Left:} (a) Redshift evolution of X-ray AGN bias in low ($M_*<10^{10.75}$\,M$_{\odot}$) and high ($M_*>10^{10.75}$\,M$_{\odot}$) mass host galaxies (blue squares and orange circles respectively). The solid lines denote the evolution of dark matter halos, as in Figure~\ref{fig:bias_z}. AGN in more massive hosts appear to reside in more massive halos. \textbf{Right:} (b) Bias of X-ray AGN in low ($M_*<10^{10.75}$\,M$_{\odot}$ and high ($M_*>10^{10.75}$\,M$_{\odot}$) mass star-forming hosts (blue pentagons and orange stars respectively). There is now no significant difference between the bias measurements of AGN in low and high mass galaxies, indicating that star-forming activity is a more important driver of clustering than stellar mass.}
\label{fig:bias_mass}
\end{figure*}

We now investigate which properties of AGN host galaxies have the dominant influence on the AGN clustering signal. To disentangle the effects of stellar mass and host galaxy spectral class on clustering, we split our AGN into high and low host galaxy stellar mass subsamples around the median mass ($10^{10.75}$ M$_{\odot}$). In Figure~\ref{fig:bias_mass}(a), we plot our clustering measurements of the AGN and find that AGN in high mass galaxies are more strongly clustered and appear to reside in halos that are almost an order of magnitude ($0.9$ dex) more massive than AGN in low mass hosts out to $z\sim2$ (at $1.7\sigma$). This is consistent with expectations from the galaxy stellar mass-halo mass relation \citep[e.g.,][]{Li_2006a, Meneux_2008, Wake_2011, Leauthaud_2012, Marulli_2013, Conselice_2018}. 

\begin{figure}
\includegraphics[width=\columnwidth,trim=0.5cm 0.0cm 1.2cm 0cm]{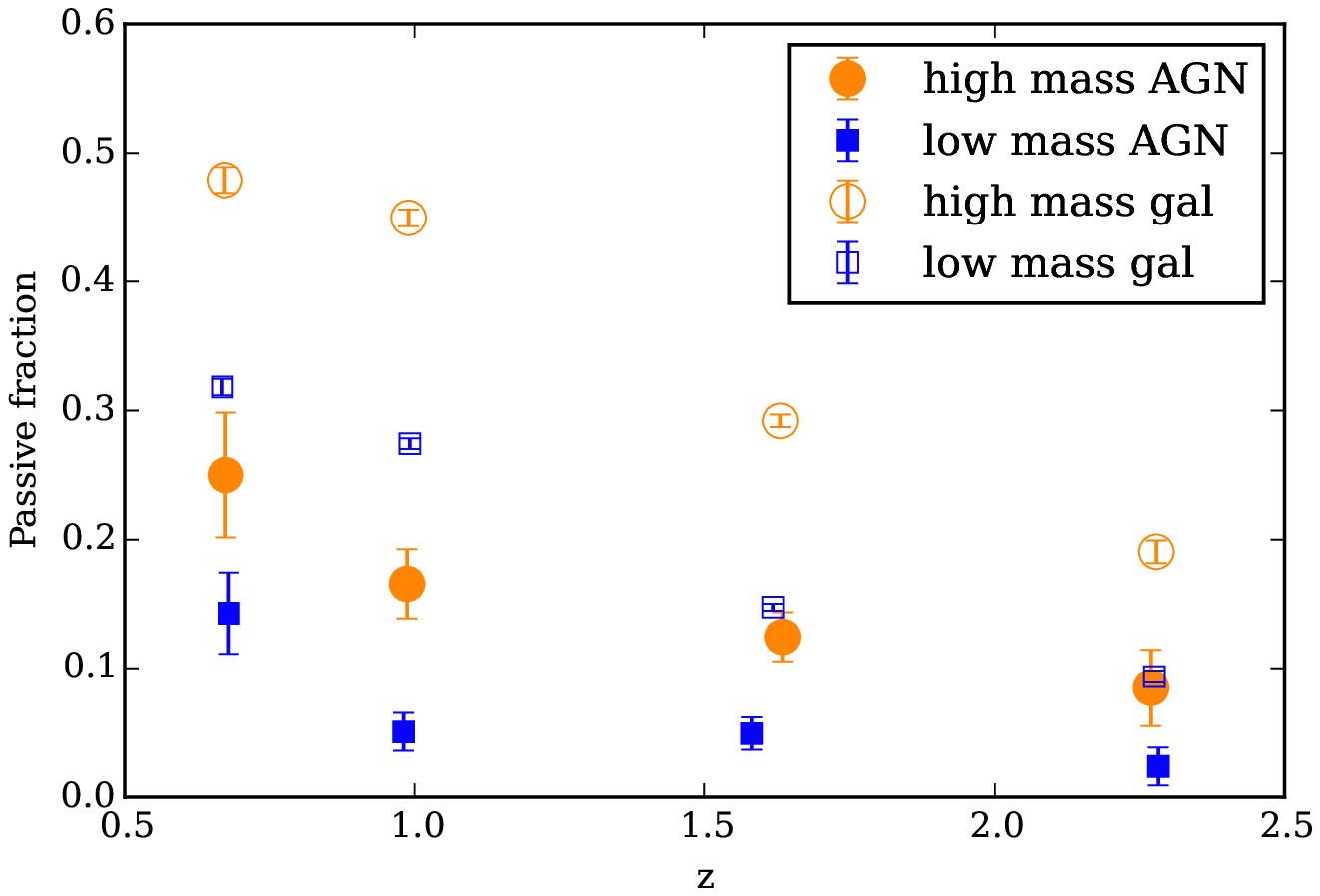}
\vspace*{-0.5cm}
\caption{Passive fractions of AGN and galaxies split by stellar mass. AGN have significantly lower passive fractions than galaxies of the same mass. AGN in high mass galaxies ($M_*>10^{10.75}$\,M$_{\odot}$; filled orange circles) have higher passive fractions than those in low mass galaxies ($M_*<10^{10.75}$\,M$_{\odot}$; filled blue squares). In general, the high mass galaxy population (open orange circles) also has higher passive fractions than the low mass counterparts (open blue squares). There is a lower passive fraction among AGN hosts than the general galaxy population, so star forming galaxies are more likely to host AGN than passive galaxies.}
\label{fig:pass_frac_mass}
\end{figure}

We next consider how host galaxy spectral class affects the clustering strength. To do this, we define the passive fraction as the fraction of a given sample that are classified as either post-starburst or red-sequence (based on super-colours). In Figure~\ref{fig:pass_frac_mass} we plot the passive fractions of AGN and non-AGN galaxies, within our two mass bins. We only select galaxies above the highest-redshift $90\%$ completeness limit of $10^{10.2}$ M$_{\odot}$ to construct the low mass galaxy sample. This plot shows that (a) the AGN detected have significantly lower passive fractions than galaxies of the same mass, and (b) the AGN in high mass hosts, like high mass galaxies, have higher passive fractions than their low mass counterparts. Since passive hosts are more clustered than star-forming hosts (see Figure~\ref{fig:bias_AGN_SF_pass_z}, \citealt{Hartley_2013}, \citealt{Coil_2017}), the difference in the measured bias between AGN in low and high mass hosts may be due to different host spectral class. There appears to be a general correlation between the difference between the passive fractions and clustering signal of AGN in high and low mass bins (e.g. the most significant difference between the passive fractions of AGN in low and high mass galaxies is at $z\sim1$, as is the most significant difference in bias of the two samples). 

To explore this further, we would ideally measure the clustering of star-forming and passive AGN separately. Although we lack the sample sizes to explore the clustering of AGN hosted by passive galaxies, we measure the clustering of AGN in star-forming galaxies split around the same median mass, and plot our result in Figure~\ref{fig:bias_mass}(b). We note that the median stellar masses in each mass/redshift bin between the two panels are consistent within error-bars ($\Delta\log$M$\leq0.04$). We find tentative evidence for a shift in the environments of AGN in star-forming host galaxies from higher halo masses at high redshift to field at low redshift. This is consistent with the effect of galaxy downsizing, where star-formation activity shifts from high mass halos to low mass halos as the Universe ages \citep{Wilkinson_2017}. Although the increasing X-ray luminosity limit with redshift implies that we are only sensitive to higher mass black holes at higher redshifts at a given Eddington ratio, we do not expect this to affect our results significantly since we see no correlation between X-ray luminosity and halo mass (see Section~\ref{sec:agn_power}). Figure~\ref{fig:bias_mass}(b) also shows that we find no difference between the dark matter halo masses of AGN in star-forming host galaxies in low and high mass bins, so the excess bias in Figure~\ref{fig:bias_mass}(a) is likely driven by the higher passive fractions in the high mass sample. We therefore find evidence that stellar mass is not the fundamental parameter that drives the excess clustering of AGN in higher mass host galaxies relative to AGN in lower mass host galaxies. Similar results have been obtained for the galaxy population, such as \citet{Coil_2017}, who find that galaxy clustering does not significantly depend on stellar mass at a given sSFR. This may imply that the stellar mass-halo mass relation is driven by the correlation between passive fraction and stellar mass. 

We caution that this analysis could be affected by the possibility that blue light from the AGN contaminates the SEDs and leads to an incorrect star-forming assignment of a passive galaxy. However, this effect is not expected to be significant as super-colour classifications are derived based on filters focussed on the rest-frame $4000$\,{\AA} break region. AGN light may also contaminate the mass measurements, but our AGN have fairly low X-ray luminosities so we do not expect this effect to have a significant impact on our conclusions (Almaini et al. in prep). In addition, \citet{Kocevski_2017} report that color contamination by lower luminosity AGN in their study is negligible, and \citet{Santini_2012} find that only $1.3\%$ of their lower-luminosity sources had a difference in their stellar mass larger than a factor of two when an AGN component is added to the stellar template. We have tested the results in Figure~\ref{fig:AGN_inact_gal} by removing the most luminous AGN ($L_\text{X} > 10^{44.4}$) from our sample, creating a new control sample, and repeating the clustering analysis. We find that the results are consistent within error-bars, and thus conclude that any contaminating AGN light does not have a major impact on our results.

\begin{table*}
\begin{tabular}{|l|c|c|c|c|c|c|}
\hline
Redshift range 	& Number of 	& Median  	  &  Fraction with 		& Median stellar mass 					& Passive host 		& Median X-ray luminosity 				\\
 		& AGN 	 	& redshift 	  &  photo-$z$ only		& ($\log \frac{M_{\rm{*}}}{\rm{M}_{\odot}}$)		& fraction		& ($\log \frac{L_{\rm{X}}}{\rm{erg\,s}^{-1}}$)		\\
\hline	
$0.5 < z < 0.8$	& 194		& 0.677		  & 0.13			& 10.8							& 0.19			& 43.2			  				\\
$0.8 < z < 1.3$	& 369		& 0.995		  & 0.21			& 10.9							& 0.11 			& 43.6			  				\\
$1.3 < z < 2.1$	& 516		& 1.599	     	  & 0.47			& 10.9							& 0.09 			& 44.1			  				\\
$2.1 < z < 2.5$	& 129		& 2.278	  	  & 0.56			& 11.0							& 0.06 			& 44.4 			  				\\
\hline
\end{tabular}
\caption{\label{tab:AGN_details} Properties of AGN samples (for which supercolour classifications are available) in each redshift bin.}
\end{table*}

\subsection{Links between clustering and AGN luminosity}
\label{sec:agn_power}

In this Section we investigate whether the clustering of AGN is dependent on the power of the AGN determined through the proxy of  X-ray luminosity. We study the correlation between the power of the black hole and the inferred dark matter halo mass by splitting our AGN into low, medium, and high X-ray luminosity bins, corresponding to $10^{43.2}\leq L_\text{x}<10^{43.8}$, $10^{43.8}\leq L_\text{x}<10^{44.4}$, and $10^{44.4}\leq L_\text{x}<10^{45.0}$ erg s$^{-1}$. We cross-check passive fractions and mass distributions between the different luminosity bins and find no significant differences between the different luminosity populations. Passive fractions vary between $5-20\%$ at all redshifts and luminosities. 

\begin{figure}
\includegraphics[width=\columnwidth,trim=0.5cm 0.0cm 1.2cm 0cm]{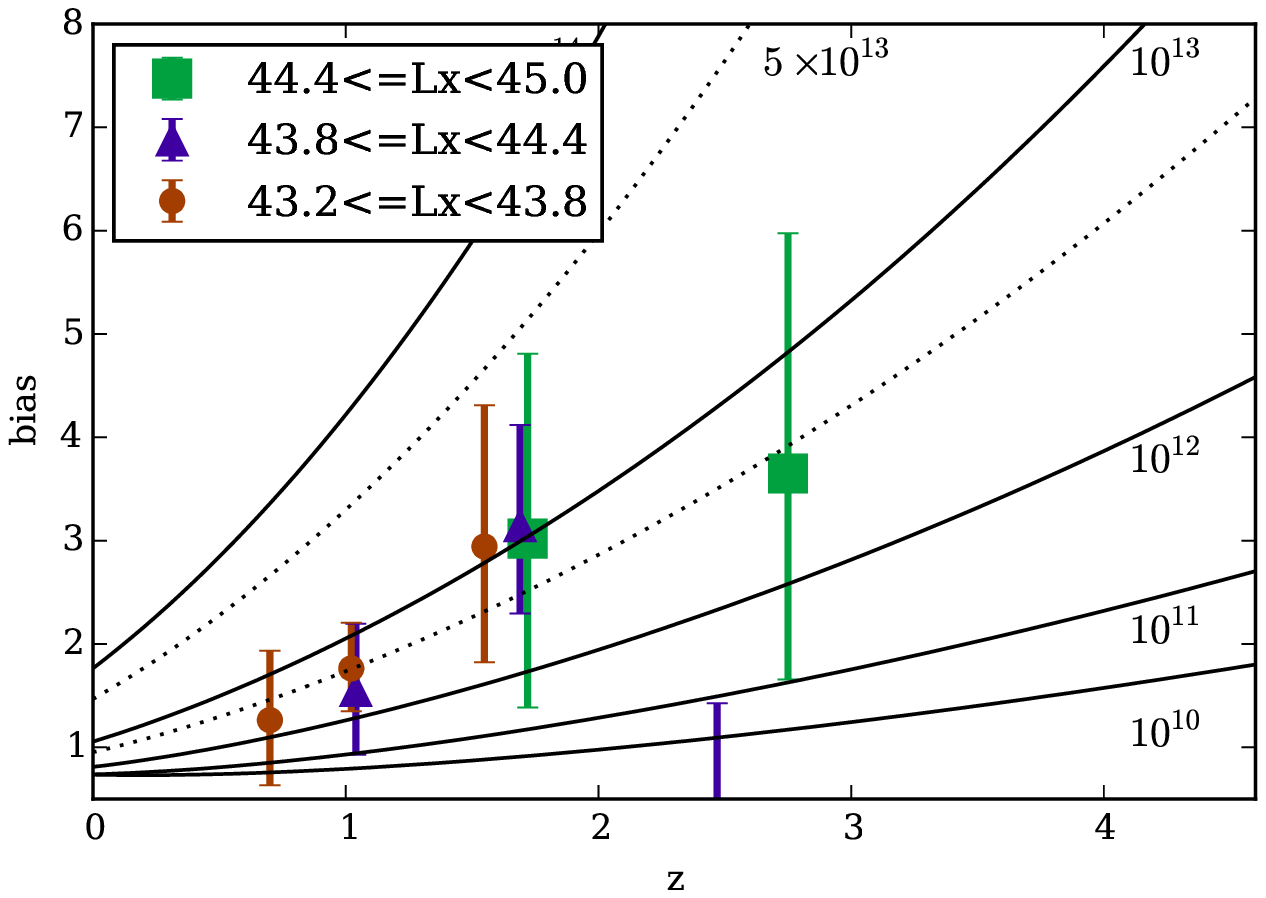}
\vspace*{-0.5cm}
\caption{Bias measurements of AGN with hard band ($2$--$10$\,keV) X-ray luminosities measured in erg s$^{-1}$ in bins of $10^{43.2}\leq L_\text{x}<10^{43.8}$, $10^{43.8}\leq L_\text{x}<10^{44.4}$, and $10^{44.4}\leq L_\text{x}<10^{45.0}$ in red circles, purple triangles, and green squares respectively. There is no correlation between redshift or luminosity and dark matter halo mass. The outlier point at $2.1<z<4.5$ is likely due to low number statistics, since it contains only $46$ AGN (the only bin with $<50$ AGN). The solid lines denote the evolution of dark matter halos, as in Figure~\ref{fig:bias_z}.}
\label{fig:bias_z_lx}
\end{figure}

While previous studies across a wide range of AGN luminosities have found similar ``group-like'' halo masses, we are able to explore this in a robust manner by splitting our AGN sample by X-ray luminosity and redshift to obtain self-consistent estimates of the bias. In Figure~\ref{fig:bias_z_lx}, we show that there is no correlation between the power of the black hole on the clustering of the AGN, which is consistent with previous results \citep[e.g.,][]{Magliocchetti_2017}. Since low, medium, and high X-ray luminosity AGN occupy similar mass halos, this implies that there may not be an environmental influence on the accretion rate of gas into the central black hole.  This also implies that the environments of AGN of all luminosities are driven by the mixed population of their hosts.

We also investigated the effects of other AGN properties such as Eddington ratio and hardness ratio, but did not have a large enough AGN sample to obtain robust results. 

\section{Discussion}
\label{sec:discussion}

We have investigated the clustering of AGN with X-ray luminosities between $43.2<$ log $L_\text{X}$ $<45.0$ and redshifts between $0.5 < z < 4.5$, and measured a bias corresponding to dark matter halos of mass $10^{12-13}$ M$_{\odot}$. Similar results have previously been interpreted to indicate a preference for AGN to reside in group-like environments  \citep[e.g.,][]{Ikeda_2015}, we have shown (see Figure~\ref{fig:AGN_inact_gal}) that the same clustering signal can be obtained from a mix of non-AGN star forming and passive galaxies that populate a range of halo masses. Based on this evidence, we suggest that AGN do not preferentially reside within a particular halo mass, and infer that AGN triggering is not primarily driven by the large-scale environment. 

We find corroborating evidence when we divide the AGN by host galaxy property. The clustering of AGN hosted by star-forming galaxies (the dominant host type) have a bias corresponding to only $10^{11-12}$ M$_{\odot}$ halos at $z < 1.5$, comparable to the bias obtained for non-AGN star-forming galaxies of similar mass and redshift. Although we do not have a sufficiently large sample to measure the clustering of AGN in passive host galaxies, we note that the clustering of this minor component of the AGN population must be significantly stronger than the AGN in star-forming hosts, because the clustering strength of the combined population average to a halo mass of $10^{12-13}$
M$_{\odot}$. Thus AGN do not preferentially reside within halos of a certain mass.

Our interpretation of the clustering signal is supported by other observational evidence, such as the lack of enhanced AGN fractions in group-like environments \citep[e.g.,][]{Shen_2007, Arnold_2009, Oh_2014, Tzanavaris_2014}.  Furthermore, if AGN preferentially reside in groups, we would expect the AGN host galaxy properties to vary with redshift since the fraction of star forming group galaxies increases with redshift \citep{Butcher_Oelmer_1978,Giodini_2012,Popesso_2012}. Instead we find that the AGN host spectral class does not vary significantly from $z=2.5$ to $z=0.5$. We therefore suggest that large-scale environment (e.g. halo mass) is not the dominant factor in triggering an AGN, although it may play a minor role. We also note that the richest and rarest cluster environments (of halo mass $\sim10^{15}$\,M$_{\odot}$) are not probed by the UDS/COSMOS fields, so we are unable to determine if AGN triggering is enhanced or suppressed in this specific regime.

Models of galaxy evolution, on the other hand, point to an important link between large-scale environment and triggering AGN activity. For instance, less massive groups have been proposed as the ideal environment for AGN activity due to an increased likelihood of mergers \citep{Hopkins_2008a, Hopkins_2008b}. Our finding is in tension with the expectation that AGN should be preferentially triggered in groups if mergers trigger AGN activity. However, we also note that observational evidence that mergers are linked to AGN triggering remains mixed \citep{Ellison_2013,Kocevski_2015,Villforth_2017,Hewlett_2017}. 

Whilst AGN do not typically reside in a special environment, we find that AGN populate special host galaxies. We have shown that the AGN hosts are not a random subset of the underlying galaxy population within the UDS and COSMOS surveys. Instead, the passive fraction of AGN is lower than the underlying galaxy population at the same stellar mass (Figure~\ref{fig:pass_frac_mass}), implying that AGN are preferentially hosted by star forming galaxies,
consistent with the findings of \citet{Aird_2017}. We infer, therefore, that the probability of triggering an AGN is correlated with some properties of the host galaxy. Consistent with this inference, \citet{Kocevski_2017} find evidence for a relationship between compactness and AGN triggering. We conclude that the triggering of AGN is likely not a simple single process, but is in fact a complex set (or sets) of conditions.

The set of conditions that trigger AGN are not likely to change drastically across $z=2.5$ to $z=0.5$ because we find that the stellar mass distribution and class of AGN host galaxies remain approximately constant in this redshift range. On the other hand, the Universe evolves drastically across this period, resulting in a significant change in galaxy properties: star formation declines \citep{Madau_Dickinson_2014}, galaxies grow in mass, and the morphological mix of galaxies evolves from predominantly irregulars to spirals and a larger fraction of early types \citep{Mortlock_2013}. The passive fraction of AGN and galaxies both increase by a factor of $2-3$ from $z=2.5$ to $z=0.5$ (see Figure~\ref{fig:pass_frac_mass}). This indicates that the fractions of star-forming and passive galaxies that host AGN do not change significantly with redshift (although we note that \citealt{Aird_2017} find evidence that quiescent galaxies are more likely to host AGN at higher redshifts).

The lack of dependence of the clustering signal on AGN power, as probed by the X-ray luminosity (Figure~\ref{fig:bias_z_lx}), suggests that the accretion rate of AGN does not have a simple dependence or correlation with its large-scale environment. This is in agreement with recent results from \citet{Yang_2018}. Furthermore, since we have shown that the clustering signal of AGN is primarily driven by the clustering properties of the AGN host population, and the mixture of host spectral classes, we infer indirectly that there is also no strong link between host spectral class and the AGN instantaneous accretion rate. 

A correlation between host spectral class and AGN accretion rate is expected because of the evidence for a correlation between the growth of stellar mass and supermassive black holes \citep{Boyle_1998, Franceschini_1999, Silverman_2008}. Given that the growth of stellar and black hole mass both rely on the availability of gas, we might expect to find higher accretion rates (through the proxy of X-ray luminosity) for star forming galaxies compared to passive galaxies of the same stellar mass. Since we find that the clustering strength of an AGN is primarily driven by its host spectral class, we would therefore expect the clustering signal of more luminous AGN to be lower than less luminous AGN, in discord with our findings.

We can explain the lack of variation in the clustering strengths for high and low luminosity AGN with the same phenomenon behind the flat SFR-$L_\text{X}$ relationship for AGN found by e.g. \citet[][]{Stanley_2015}. \citet{Hickox_2014} proposed that large X-ray AGN variability on short timescales (relative to that of star-formation) dilutes the intrinsic correlation between SFR and $L_\text{X}$. To reproduce the underlying relation we must average over the most variable quantity ($L_\text{X}$ in this case). High X-ray variability on short-timescales could also dilute any intrinsic $L_\text{X}$-host spectral class correlation, which would then dilute any variation in the clustering of AGN of different power. We note that a similar result may also be obtained if star-formation and black hole accretion are not coeval \citep{Wild_2010}.

Drawing our interpretations together, the triggering of AGN activity likely depends on a complex set of conditions. We find a lack of correlation between AGN power and clustering signal, and that a mix of non-AGN star-forming and passive galaxies can reproduce the same clustering signal as AGN. Together this suggests that large scale environment plays at most a minor role in triggering AGN activity. Availability of fuel may be inferred to play a major role since star-forming galaxies are more likely to host AGN at all epochs. This is supported by the well-known increased prevalence of high luminosity AGN at high redshift compared to low redshift (e.g. Figure~\ref{fig:bias_z_lx}). However, environment may be a relevant factor if mergers only trigger the most luminous AGN, as merger rates decline as the Universe ages \citep{Conselice_2008}. Furthermore, disks and bars may trigger nuclear activity \citep[e.g.,][]{Knapen_2000} as disk instabilities have been proposed to enhance gas flow to the nuclei of galaxies \citep{Dekel_2009}. Therefore, multiple mechanisms could be at play in the triggering of AGN, unlikely to be defined by solely large-scale environment or availability of fuel.

We note that our work is limited to X-ray selection, which tends to select more powerful and rapidly accreting AGN. Selection effects may play an important role in the interpretations and conclusions from AGN clustering results. Previous studies have found that X-ray and radio AGN are more clustered than mid-IR-selected AGN \citep[e.g.,][]{Hickox_2009, Mendez_2016}. However, when each AGN sample was studied using galaxy samples matched in stellar mass, star-formation rate, and redshift, \citet{Mendez_2016} find no significant differences between the clustering properties of the AGN samples. They find that AGN selected in different wavelengths appear to have different clustering properties simply because they are sampling different host populations with different stellar mass and SFR distributions. Therefore, the method of AGN selection introduces inherent biases in the host galaxy properties and likely determines the clustering signal. Our study could be repeated using optical, near-infrared, and radio selected AGN with samples matched in mass, (s)SFR, and redshift to obtain a clearer picture of the impact of host galaxies on AGN clustering measurements. The implication for future AGN clustering studies is that samples must be divided by host galaxy properties as the clustering signal from AGN likely represents that of host galaxies.

To summarise our interpretations, the triggering of AGN activity likely depends on a complex set of conditions that do not depend solely on large-scale environment or availability of fuel. The parameters that define the triggering of AGN could have important implications for our understanding of galaxy evolution, particularly as our most sophisticated galaxy evolution models invoke AGN feedback as a key ingredient in reproducing the stellar mass functions of galaxies.

\section{Conclusions}

In this work we study the clustering properties of a flux-limited sample of hard X-ray selected AGN from $z\sim4.5$ to $z\sim0.5$, using the COSMOS and UDS multi-wavelength surveys. We compare them to a control galaxy sample designed to have similar distributions of stellar mass, spectral class, and redshift. We investigate the role of properties of host galaxies (e.g. stellar mass) and of the central AGN (X-ray luminosity) as a function of redshift. We find that the clustering properties of AGN cannot be naively linked to large-scale environments as host galaxy properties play a significant role in the clustering measurements. To summarise our findings:

\begin{enumerate}

\item We find that hard X-ray selected AGN in the UDS and COSMOS fields have bias parameters that correspond to a typical halo mass of $10^{12-13}$\,M$_{\odot}$.  No evidence for evolution in halo mass is found between $0.5<z<4.5$. 

\item We compare the clustering of AGN to star-forming and passive galaxy populations matched in mass and redshift distributions. We find that the clustering of AGN lies in between the mass-matched star-forming and passive populations. 

\item We can reproduce the clustering signal of AGN with an inactive galaxy population closely matched in spectral class, mass, and redshift distributions as the AGN host galaxies. We thus find that the mixed population of star-forming and passive AGN host galaxies drives the clustering properties of AGN.

\item We split AGN by host galaxy stellar mass and find an excess clustering in the high mass sample. The stellar mass dependence of clustering disappears once passive galaxies are removed from the samples, as we find no difference in their clustering properties. Therefore, we conclude that AGN clustering depends more strongly on the spectral class of the host galaxies than stellar mass.

\item We find no difference in the clustering properties of low, medium and high X-ray luminosity AGN. The triggering of AGN activity is likely determined by a complex set of conditions that do not depend solely on large-scale environment.

\end{enumerate}

\section*{Acknowledgements}

We thank the anonymous referee for their insightful comments, which improved the paper. CK is supported by a Vice-Chancellor's Scholarship for Research Excellence. We thank Will Hartley for useful discussions. This work is based on observations made with ESO Telescopes at the La Silla Paranal Observatory under programme ID 089.A-0126. This work also uses data from ESO telescopes at the Paranal Observatory (programmes 094.A-0410 and 180.A-0776; PI: Almaini). We are grateful to the staff at UKIRT for their tireless efforts in ensuring the success of the UDS project. We also wish to recognize and acknowledge the very significant cultural role and reverence that the summit of Mauna Kea has within the indigenous Hawaiian community. We were most fortunate to have the opportunity to conduct observations from this mountain.

\bibliographystyle{mnras}
\bibliography{biblio4} 

\begin{thebibliography}{}
\makeatletter
\relax
\def\mn@urlcharsother{\let\do\@makeother \do\$\do\&\do\#\do\^\do\_\do\%\do\~}
\def\mn@doi{\begingroup\mn@urlcharsother \@ifnextchar [ {\mn@doi@}
  {\mn@doi@[]}}
\def\mn@doi@[#1]#2{\def\@tempa{#1}\ifx\@tempa\@empty \href
  {http://dx.doi.org/#2} {doi:#2}\else \href {http://dx.doi.org/#2} {#1}\fi
  \endgroup}
\def\mn@eprint#1#2{\mn@eprint@#1:#2::\@nil}
\def\mn@eprint@arXiv#1{\href {http://arxiv.org/abs/#1} {{\tt arXiv:#1}}}
\def\mn@eprint@dblp#1{\href {http://dblp.uni-trier.de/rec/bibtex/#1.xml}
  {dblp:#1}}
\def\mn@eprint@#1:#2:#3:#4\@nil{\def\@tempa {#1}\def\@tempb {#2}\def\@tempc
  {#3}\ifx \@tempc \@empty \let \@tempc \@tempb \let \@tempb \@tempa \fi \ifx
  \@tempb \@empty \def\@tempb {arXiv}\fi \@ifundefined
  {mn@eprint@\@tempb}{\@tempb:\@tempc}{\expandafter \expandafter \csname
  mn@eprint@\@tempb\endcsname \expandafter{\@tempc}}}

\bibitem[\protect\citeauthoryear{{Aird}, {Coil}  \& {Georgakakis}}{{Aird}
  et~al.}{2017}]{Aird_2017}
{Aird} J.,  {Coil} A.~L.,   {Georgakakis} A.,  2017, \mn@doi [\mnras]
  {10.1093/mnras/stw2932}, 465, 3390

\bibitem[\protect\citeauthoryear{{Akiyama} et~al.,}{{Akiyama}
  et~al.}{2015}]{Akiyama_2015}
{Akiyama} M.,  et~al., 2015, \mn@doi [\pasj] {10.1093/pasj/psv050}, 67, 82

\bibitem[\protect\citeauthoryear{{Allevato} et~al.,}{{Allevato}
  et~al.}{2011}]{Allevato_2011}
{Allevato} V.,  et~al., 2011, \mn@doi [\apj] {10.1088/0004-637X/736/2/99}, 736,
  99

\bibitem[\protect\citeauthoryear{{Allevato} et~al.,}{{Allevato}
  et~al.}{2016}]{Allevato_2016}
{Allevato} V.,  et~al., 2016, \mn@doi [\apj] {10.3847/0004-637X/832/1/70}, 832,
  70

\bibitem[\protect\citeauthoryear{Almaini et~al.,}{Almaini
  et~al.}{2017}]{Almaini_2017}
Almaini O.,  et~al., 2017, \mn@doi [MNRAS] {10.1093/mnras/stx1957}, 472, 1401

\bibitem[\protect\citeauthoryear{{Arnold}, {Martini}, {Mulchaey}, {Berti}  \&
  {Jeltema}}{{Arnold} et~al.}{2009}]{Arnold_2009}
{Arnold} T.~J.,  {Martini} P.,  {Mulchaey} J.~S.,  {Berti} A.,   {Jeltema}
  T.~E.,  2009, \mn@doi [\apj] {10.1088/0004-637X/707/2/1691}, 707, 1691

\bibitem[\protect\citeauthoryear{{Ashby} et~al.,}{{Ashby}
  et~al.}{2013}]{Ashby_2013}
{Ashby} M.~L.~N.,  et~al., 2013, \mn@doi [\apj] {10.1088/0004-637X/769/1/80},
  769, 80

\bibitem[\protect\citeauthoryear{{Benson}, {Cole}, {Frenk}, {Baugh}  \&
  {Lacey}}{{Benson} et~al.}{2000}]{Benson_2000}
{Benson} A.~J.,  {Cole} S.,  {Frenk} C.~S.,  {Baugh} C.~M.,   {Lacey} C.~G.,
  2000, \mn@doi [\mnras] {10.1046/j.1365-8711.2000.03101.x}, 311, 793

\bibitem[\protect\citeauthoryear{{Boyle}, {Georgantopoulos}, {Blair},
  {Stewart}, {Griffiths}, {Shanks}, {Gunn}  \& {Almaini}}{{Boyle}
  et~al.}{1998}]{Boyle_1998}
{Boyle} B.~J.,  {Georgantopoulos} I.,  {Blair} A.~J.,  {Stewart} G.~C.,
  {Griffiths} R.~E.,  {Shanks} T.,  {Gunn} K.~F.,   {Almaini} O.,  1998,
  \mn@doi [\mnras] {10.1046/j.1365-8711.1998.01098.x}, 296, 1

\bibitem[\protect\citeauthoryear{{Bradshaw} et~al.,}{{Bradshaw}
  et~al.}{2011}]{Bradshaw_2011}
{Bradshaw} E.~J.,  et~al., 2011, \mn@doi [\mnras]
  {10.1111/j.1365-2966.2011.18888.x}, 415, 2626

\bibitem[\protect\citeauthoryear{{Bruzual} \& {Charlot}}{{Bruzual} \&
  {Charlot}}{2003}]{Bruzual_Charlot_2003}
{Bruzual} G.,  {Charlot} S.,  2003, \mn@doi [\mnras]
  {10.1046/j.1365-8711.2003.06897.x}, 344, 1000

\bibitem[\protect\citeauthoryear{{Butcher} \& {Oemler}}{{Butcher} \&
  {Oemler}}{1978}]{Butcher_Oelmer_1978}
{Butcher} H.,  {Oemler} Jr. A.,  1978, \mn@doi [\apj] {10.1086/155751}, 219, 18

\bibitem[\protect\citeauthoryear{{Chabrier}}{{Chabrier}}{2003}]{Chabrier_2003}
{Chabrier} G.,  2003, \mn@doi [\pasp] {10.1086/376392}, 115, 763

\bibitem[\protect\citeauthoryear{{Civano} et~al.,}{{Civano}
  et~al.}{2012}]{Civano_2012}
{Civano} F.,  et~al., 2012, \mn@doi [\apjs] {10.1088/0067-0049/201/2/30}, 201,
  30

\bibitem[\protect\citeauthoryear{{Civano} et~al.,}{{Civano}
  et~al.}{2016}]{Civano_2016}
{Civano} F.,  et~al., 2016, \mn@doi [\apj] {10.3847/0004-637X/819/1/62}, 819,
  62

\bibitem[\protect\citeauthoryear{{Coil}}{{Coil}}{2013}]{Coil_2013}
{Coil} A.~L.,  2013, {The Large-Scale Structure of the Universe}.
p.~387, \mn@doi{10.1007/978-94-007-5609-0_8}

\bibitem[\protect\citeauthoryear{{Coil}, {Mendez}, {Eisenstein}  \&
  {Moustakas}}{{Coil} et~al.}{2017}]{Coil_2017}
{Coil} A.~L.,  {Mendez} A.~J.,  {Eisenstein} D.~J.,   {Moustakas} J.,  2017,
  \mn@doi [\apj] {10.3847/1538-4357/aa63ec}, 838, 87

\bibitem[\protect\citeauthoryear{{Conselice}}{{Conselice}}{2014}]{Conselice_2014}
{Conselice} C.~J.,  2014, \mn@doi [\araa]
  {10.1146/annurev-astro-081913-040037}, 52, 291

\bibitem[\protect\citeauthoryear{{Conselice}, {Rajgor}  \& {Myers}}{{Conselice}
  et~al.}{2008}]{Conselice_2008}
{Conselice} C.~J.,  {Rajgor} S.,   {Myers} R.,  2008, \mn@doi [\mnras]
  {10.1111/j.1365-2966.2008.13069.x}, 386, 909

\bibitem[\protect\citeauthoryear{{Conselice}, {Twite}, {Palamara}  \&
  {Hartley}}{{Conselice} et~al.}{2018}]{Conselice_2018}
{Conselice} C.~J.,  {Twite} J.~W.,  {Palamara} D.~P.,   {Hartley} W.,  2018,
  \mn@doi [\apj] {10.3847/1538-4357/aacda8}, 863, 42

\bibitem[\protect\citeauthoryear{{Croom} et~al.,}{{Croom}
  et~al.}{2005}]{Croom_2005}
{Croom} S.~M.,  et~al., 2005, \mn@doi [\mnras]
  {10.1111/j.1365-2966.2004.08379.x}, 356, 415

\bibitem[\protect\citeauthoryear{{De Lucia}, {Springel}, {White}, {Croton}  \&
  {Kauffmann}}{{De Lucia} et~al.}{2006}]{DeLucia_2006}
{De Lucia} G.,  {Springel} V.,  {White} S.~D.~M.,  {Croton} D.,   {Kauffmann}
  G.,  2006, \mn@doi [\mnras] {10.1111/j.1365-2966.2005.09879.x}, 366, 499

\bibitem[\protect\citeauthoryear{{Dekel}, {Sari}  \& {Ceverino}}{{Dekel}
  et~al.}{2009}]{Dekel_2009}
{Dekel} A.,  {Sari} R.,   {Ceverino} D.,  2009, \mn@doi [\apj]
  {10.1088/0004-637X/703/1/785}, 703, 785

\bibitem[\protect\citeauthoryear{{Digby-North} et~al.,}{{Digby-North}
  et~al.}{2010}]{Digby-North_2010}
{Digby-North} J.~A.,  et~al., 2010, \mn@doi [\mnras]
  {10.1111/j.1365-2966.2010.16977.x}, 407, 846

\bibitem[\protect\citeauthoryear{{Dressler}, {Smail}, {Poggianti}, {Butcher},
  {Couch}, {Ellis}  \& {Oemler}}{{Dressler} et~al.}{1999}]{Dressler_1999}
{Dressler} A.,  {Smail} I.,  {Poggianti} B.~M.,  {Butcher} H.,  {Couch} W.~J.,
  {Ellis} R.~S.,   {Oemler} Jr. A.,  1999, \mn@doi [\apjs] {10.1086/313213},
  122, 51

\bibitem[\protect\citeauthoryear{{Eftekharzadeh} et~al.,}{{Eftekharzadeh}
  et~al.}{2015}]{Eftekharzadeh_2015}
{Eftekharzadeh} S.,  et~al., 2015, \mn@doi [\mnras] {10.1093/mnras/stv1763},
  453, 2779

\bibitem[\protect\citeauthoryear{{Ellison}, {Mendel}, {Scudder}, {Patton}  \&
  {Palmer}}{{Ellison} et~al.}{2013}]{Ellison_2013}
{Ellison} S.~L.,  {Mendel} J.~T.,  {Scudder} J.~M.,  {Patton} D.~R.,   {Palmer}
  M.~J.~D.,  2013, \mn@doi [\mnras] {10.1093/mnras/sts546}, 430, 3128

\bibitem[\protect\citeauthoryear{{Franceschini}, {Hasinger}, {Miyaji}  \&
  {Malquori}}{{Franceschini} et~al.}{1999}]{Franceschini_1999}
{Franceschini} A.,  {Hasinger} G.,  {Miyaji} T.,   {Malquori} D.,  1999,
  \mn@doi [\mnras] {10.1046/j.1365-8711.1999.03078.x}, 310, L5

\bibitem[\protect\citeauthoryear{{Furusawa} et~al.,}{{Furusawa}
  et~al.}{2008}]{Furusawa_2008}
{Furusawa} H.,  et~al., 2008, \mn@doi [\apjs] {10.1086/527321}, 176, 1

\bibitem[\protect\citeauthoryear{{Garc{\'{\i}}a-Vergara}, {Hennawi},
  {Barrientos}  \& {Rix}}{{Garc{\'{\i}}a-Vergara}
  et~al.}{2017}]{Garcia-Vergara_2017}
{Garc{\'{\i}}a-Vergara} C.,  {Hennawi} J.~F.,  {Barrientos} L.~F.,   {Rix}
  H.-W.,  2017, \mn@doi [\apj] {10.3847/1538-4357/aa8b69}, 848, 7

\bibitem[\protect\citeauthoryear{{Garmire}, {Bautz}, {Ford}, {Nousek}  \&
  {Ricker}}{{Garmire} et~al.}{2003}]{Garmire_2003}
{Garmire} G.~P.,  {Bautz} M.~W.,  {Ford} P.~G.,  {Nousek} J.~A.,   {Ricker} Jr.
  G.~R.,  2003, in {Truemper} J.~E.,  {Tananbaum} H.~D.,  eds,  \procspie Vol.
  4851, X-Ray and Gamma-Ray Telescopes and Instruments for Astronomy.. pp
  28--44, \mn@doi{10.1117/12.461599}

\bibitem[\protect\citeauthoryear{{Giodini} et~al.,}{{Giodini}
  et~al.}{2012}]{Giodini_2012}
{Giodini} S.,  et~al., 2012, \mn@doi [\aap] {10.1051/0004-6361/201117696}, 538,
  A104

\bibitem[\protect\citeauthoryear{{Gisler}}{{Gisler}}{1978}]{Gisler_1978}
{Gisler} G.~R.,  1978, \mn@doi [\mnras] {10.1093/mnras/183.4.633}, 183, 633

\bibitem[\protect\citeauthoryear{{Hartley} et~al.,}{{Hartley}
  et~al.}{2013}]{Hartley_2013}
{Hartley} W.~G.,  et~al., 2013, \mn@doi [\mnras] {10.1093/mnras/stt383}, 431,
  3045

\bibitem[\protect\citeauthoryear{{Hewlett}, {Villforth}, {Wild},
  {Mendez-Abreu}, {Pawlik}  \& {Rowlands}}{{Hewlett}
  et~al.}{2017}]{Hewlett_2017}
{Hewlett} T.,  {Villforth} C.,  {Wild} V.,  {Mendez-Abreu} J.,  {Pawlik} M.,
  {Rowlands} K.,  2017, \mn@doi [\mnras] {10.1093/mnras/stx997}, 470, 755

\bibitem[\protect\citeauthoryear{{Hickox} et~al.,}{{Hickox}
  et~al.}{2009}]{Hickox_2009}
{Hickox} R.~C.,  et~al., 2009, \mn@doi [\apj] {10.1088/0004-637X/696/1/891},
  696, 891

\bibitem[\protect\citeauthoryear{{Hickox}, {Mullaney}, {Alexander}, {Chen},
  {Civano}, {Goulding}  \& {Hainline}}{{Hickox} et~al.}{2014}]{Hickox_2014}
{Hickox} R.~C.,  {Mullaney} J.~R.,  {Alexander} D.~M.,  {Chen} C.-T.~J.,
  {Civano} F.~M.,  {Goulding} A.~D.,   {Hainline} K.~N.,  2014, \mn@doi [\apj]
  {10.1088/0004-637X/782/1/9}, 782, 9

\bibitem[\protect\citeauthoryear{{Hinshaw} et~al.,}{{Hinshaw}
  et~al.}{2013}]{Hinshaw_2013}
{Hinshaw} G.,  et~al., 2013, \mn@doi [\apjs] {10.1088/0067-0049/208/2/19}, 208,
  19

\bibitem[\protect\citeauthoryear{{Hopkins}, {Hernquist}, {Cox}  \& {Kere{\v
  s}}}{{Hopkins} et~al.}{2008a}]{Hopkins_2008a}
{Hopkins} P.~F.,  {Hernquist} L.,  {Cox} T.~J.,   {Kere{\v s}} D.,  2008a,
  \mn@doi [\apjs] {10.1086/524362}, 175, 356

\bibitem[\protect\citeauthoryear{{Hopkins}, {Cox}, {Kere{\v s}}  \&
  {Hernquist}}{{Hopkins} et~al.}{2008b}]{Hopkins_2008b}
{Hopkins} P.~F.,  {Cox} T.~J.,  {Kere{\v s}} D.,   {Hernquist} L.,  2008b,
  \mn@doi [\apjs] {10.1086/524363}, 175, 390

\bibitem[\protect\citeauthoryear{{Ikeda} et~al.,}{{Ikeda}
  et~al.}{2015}]{Ikeda_2015}
{Ikeda} H.,  et~al., 2015, \mn@doi [\apj] {10.1088/0004-637X/809/2/138}, 809,
  138

\bibitem[\protect\citeauthoryear{{Jarvis} et~al.,}{{Jarvis}
  et~al.}{2013}]{Jarvis_2013}
{Jarvis} M.~J.,  et~al., 2013, \mn@doi [\mnras] {10.1093/mnras/sts118}, 428,
  1281

\bibitem[\protect\citeauthoryear{{Kauffmann}, {White}, {Heckman}, {M{\'e}nard},
  {Brinchmann}, {Charlot}, {Tremonti}  \& {Brinkmann}}{{Kauffmann}
  et~al.}{2004}]{Kauffmann_2004}
{Kauffmann} G.,  {White} S.~D.~M.,  {Heckman} T.~M.,  {M{\'e}nard} B.,
  {Brinchmann} J.,  {Charlot} S.,  {Tremonti} C.,   {Brinkmann} J.,  2004,
  \mn@doi [\mnras] {10.1111/j.1365-2966.2004.08117.x}, 353, 713

\bibitem[\protect\citeauthoryear{{Knapen}, {Shlosman}  \& {Peletier}}{{Knapen}
  et~al.}{2000}]{Knapen_2000}
{Knapen} J.~H.,  {Shlosman} I.,   {Peletier} R.~F.,  2000, \mn@doi [\apj]
  {10.1086/308266}, 529, 93

\bibitem[\protect\citeauthoryear{{Kocevski} et~al.,}{{Kocevski}
  et~al.}{2015}]{Kocevski_2015}
{Kocevski} D.~D.,  et~al., 2015, \mn@doi [\apj] {10.1088/0004-637X/814/2/104},
  814, 104

\bibitem[\protect\citeauthoryear{{Kocevski} et~al.,}{{Kocevski}
  et~al.}{2017}]{Kocevski_2017}
{Kocevski} D.~D.,  et~al., 2017, \mn@doi [\apj] {10.3847/1538-4357/aa8566},
  846, 112

\bibitem[\protect\citeauthoryear{{Kocevski} et~al.,}{{Kocevski}
  et~al.}{2018}]{Kocevski_2018}
{Kocevski} D.~D.,  et~al., 2018, \mn@doi [\apjs] {10.3847/1538-4365/aab9b4},
  236, 48

\bibitem[\protect\citeauthoryear{{Kormendy} \& {Ho}}{{Kormendy} \&
  {Ho}}{2013}]{Kormendy_Ho_2013}
{Kormendy} J.,  {Ho} L.~C.,  2013, \mn@doi [\araa]
  {10.1146/annurev-astro-082708-101811}, 51, 511

\bibitem[\protect\citeauthoryear{{Koutoulidis}, {Plionis}, {Georgantopoulos}
  \& {Fanidakis}}{{Koutoulidis} et~al.}{2013}]{Koutoulidis_2013}
{Koutoulidis} L.,  {Plionis} M.,  {Georgantopoulos} I.,   {Fanidakis} N.,
  2013, \mn@doi [\mnras] {10.1093/mnras/sts119}, 428, 1382

\bibitem[\protect\citeauthoryear{{Krishnan} et~al.,}{{Krishnan}
  et~al.}{2017}]{Krishnan_2017}
{Krishnan} C.,  et~al., 2017, \mn@doi [\mnras] {10.1093/mnras/stx1315}, 470,
  2170

\bibitem[\protect\citeauthoryear{{Krumpe}, {Miyaji}, {Coil}  \&
  {Aceves}}{{Krumpe} et~al.}{2012}]{Krumpe_2012}
{Krumpe} M.,  {Miyaji} T.,  {Coil} A.~L.,   {Aceves} H.,  2012, \mn@doi [\apj]
  {10.1088/0004-637X/746/1/1}, 746, 1

\bibitem[\protect\citeauthoryear{{Laigle} et~al.,}{{Laigle}
  et~al.}{2016}]{Laigle_2016}
{Laigle} C.,  et~al., 2016, \mn@doi [\apjs] {10.3847/0067-0049/224/2/24}, 224,
  24

\bibitem[\protect\citeauthoryear{{Landy} \& {Szalay}}{{Landy} \&
  {Szalay}}{1993}]{Landy_Szalay_1993}
{Landy} S.~D.,  {Szalay} A.~S.,  1993, \mn@doi [\apj] {10.1086/172900}, 412, 64

\bibitem[\protect\citeauthoryear{{Lawrence} et~al.,}{{Lawrence}
  et~al.}{2007}]{Lawrence_2007}
{Lawrence} A.,  et~al., 2007, \mn@doi [\mnras]
  {10.1111/j.1365-2966.2007.12040.x}, 379, 1599

\bibitem[\protect\citeauthoryear{{Leauthaud} et~al.,}{{Leauthaud}
  et~al.}{2012}]{Leauthaud_2012}
{Leauthaud} A.,  et~al., 2012, \mn@doi [\apj] {10.1088/0004-637X/744/2/159},
  744, 159

\bibitem[\protect\citeauthoryear{{Lehmer} et~al.,}{{Lehmer}
  et~al.}{2009}]{Lehmer_2009}
{Lehmer} B.~D.,  et~al., 2009, \mn@doi [\apj] {10.1088/0004-637X/691/1/687},
  691, 687

\bibitem[\protect\citeauthoryear{{Li}, {Kauffmann}, {Jing}, {White},
  {B{\"o}rner}  \& {Cheng}}{{Li} et~al.}{2006}]{Li_2006a}
{Li} C.,  {Kauffmann} G.,  {Jing} Y.~P.,  {White} S.~D.~M.,  {B{\"o}rner} G.,
  {Cheng} F.~Z.,  2006, \mn@doi [\mnras] {10.1111/j.1365-2966.2006.10066.x},
  368, 21

\bibitem[\protect\citeauthoryear{{Limber}}{{Limber}}{1954}]{Limber_1954}
{Limber} D.~N.,  1954, \mn@doi [\apj] {10.1086/145870}, 119, 655

\bibitem[\protect\citeauthoryear{{Madau} \& {Dickinson}}{{Madau} \&
  {Dickinson}}{2014}]{Madau_Dickinson_2014}
{Madau} P.,  {Dickinson} M.,  2014, \mn@doi [\araa]
  {10.1146/annurev-astro-081811-125615}, 52, 415

\bibitem[\protect\citeauthoryear{{Magliocchetti} \& {Maddox}}{{Magliocchetti}
  \& {Maddox}}{1999}]{Magliocchetti_Maddox_1999}
{Magliocchetti} M.,  {Maddox} S.~J.,  1999, \mn@doi [\mnras]
  {10.1046/j.1365-8711.1999.02612.x}, 306, 988

\bibitem[\protect\citeauthoryear{{Magliocchetti}, {Popesso}, {Brusa},
  {Salvato}, {Laigle}, {McCracken}  \& {Ilbert}}{{Magliocchetti}
  et~al.}{2017}]{Magliocchetti_2017}
{Magliocchetti} M.,  {Popesso} P.,  {Brusa} M.,  {Salvato} M.,  {Laigle} C.,
  {McCracken} H.~J.,   {Ilbert} O.,  2017, \mn@doi [\mnras]
  {10.1093/mnras/stw2541}, 464, 3271

\bibitem[\protect\citeauthoryear{{Marchesi} et~al.,}{{Marchesi}
  et~al.}{2016}]{Marchesi_2016}
{Marchesi} S.,  et~al., 2016, \mn@doi [\apj] {10.3847/0004-637X/817/1/34}, 817,
  34

\bibitem[\protect\citeauthoryear{{Marulli} et~al.,}{{Marulli}
  et~al.}{2013}]{Marulli_2013}
{Marulli} F.,  et~al., 2013, \mn@doi [\aap] {10.1051/0004-6361/201321476}, 557,
  A17

\bibitem[\protect\citeauthoryear{{McCracken} et~al.,}{{McCracken}
  et~al.}{2012}]{McCracken_2012}
{McCracken} H.~J.,  et~al., 2012, \mn@doi [\aap] {10.1051/0004-6361/201219507},
  544, A156

\bibitem[\protect\citeauthoryear{{Mendez} et~al.,}{{Mendez}
  et~al.}{2016}]{Mendez_2016}
{Mendez} A.~J.,  et~al., 2016, \mn@doi [\apj] {10.3847/0004-637X/821/1/55},
  821, 55

\bibitem[\protect\citeauthoryear{{Meneux} et~al.,}{{Meneux}
  et~al.}{2008}]{Meneux_2008}
{Meneux} B.,  et~al., 2008, \mn@doi [\aap] {10.1051/0004-6361:20078182}, 478,
  299

\bibitem[\protect\citeauthoryear{{Miyaji}, {Krumpe}, {Coil}  \&
  {Aceves}}{{Miyaji} et~al.}{2011}]{Miyaji_2011}
{Miyaji} T.,  {Krumpe} M.,  {Coil} A.~L.,   {Aceves} H.,  2011, \mn@doi [\apj]
  {10.1088/0004-637X/726/2/83}, 726, 83

\bibitem[\protect\citeauthoryear{{Mo} \& {White}}{{Mo} \&
  {White}}{2002}]{Mo_White_2002}
{Mo} H.~J.,  {White} S.~D.~M.,  2002, \mn@doi [\mnras]
  {10.1046/j.1365-8711.2002.05723.x}, 336, 112

\bibitem[\protect\citeauthoryear{{Mortlock} et~al.,}{{Mortlock}
  et~al.}{2013}]{Mortlock_2013}
{Mortlock} A.,  et~al., 2013, \mn@doi [\mnras] {10.1093/mnras/stt793}, 433,
  1185

\bibitem[\protect\citeauthoryear{{Mountrichas} \& {Georgakakis}}{{Mountrichas}
  \& {Georgakakis}}{2012}]{Mountrichas_Georgakakis_2012}
{Mountrichas} G.,  {Georgakakis} A.,  2012, \mn@doi [\mnras]
  {10.1111/j.1365-2966.2011.20059.x}, 420, 514

\bibitem[\protect\citeauthoryear{{Oh} et~al.,}{{Oh} et~al.}{2014}]{Oh_2014}
{Oh} S.,  et~al., 2014, \mn@doi [\apj] {10.1088/0004-637X/790/1/43}, 790, 43

\bibitem[\protect\citeauthoryear{{Peebles}}{{Peebles}}{1980}]{Peebles_1980}
{Peebles} P.~J.~E.,  1980, {The large-scale structure of the universe}

\bibitem[\protect\citeauthoryear{{Polletta} et~al.,}{{Polletta}
  et~al.}{2007}]{Polletta_2007}
{Polletta} M.,  et~al., 2007, \mn@doi [\apj] {10.1086/518113}, 663, 81

\bibitem[\protect\citeauthoryear{{Popesso} et~al.,}{{Popesso}
  et~al.}{2012}]{Popesso_2012}
{Popesso} P.,  et~al., 2012, \mn@doi [\aap] {10.1051/0004-6361/201117973}, 537,
  A58

\bibitem[\protect\citeauthoryear{{Powell} et~al.,}{{Powell}
  et~al.}{2018}]{Powell_2018}
{Powell} M.~C.,  et~al., 2018, \mn@doi [\apj] {10.3847/1538-4357/aabd7f}, 858,
  110

\bibitem[\protect\citeauthoryear{{Pozzetti} et~al.,}{{Pozzetti}
  et~al.}{2010}]{Pozzetti_2010}
{Pozzetti} L.,  et~al., 2010, \mn@doi [\aap] {10.1051/0004-6361/200913020},
  523, A13

\bibitem[\protect\citeauthoryear{{Roche} \& {Eales}}{{Roche} \&
  {Eales}}{1999}]{Roche_Eales_1999}
{Roche} N.,  {Eales} S.~A.,  1999, \mn@doi [\mnras]
  {10.1046/j.1365-8711.1999.02652.x}, 307, 703

\bibitem[\protect\citeauthoryear{{Ross} et~al.,}{{Ross}
  et~al.}{2009}]{Ross_2009}
{Ross} N.~P.,  et~al., 2009, \mn@doi [\apj] {10.1088/0004-637X/697/2/1634},
  697, 1634

\bibitem[\protect\citeauthoryear{{Santini} et~al.,}{{Santini}
  et~al.}{2012}]{Santini_2012}
{Santini} P.,  et~al., 2012, \mn@doi [\aap] {10.1051/0004-6361/201118266}, 540,
  A109

\bibitem[\protect\citeauthoryear{{Scoville} et~al.,}{{Scoville}
  et~al.}{2007}]{Scoville_2007}
{Scoville} N.,  et~al., 2007, \mn@doi [\apjs] {10.1086/516585}, 172, 1

\bibitem[\protect\citeauthoryear{{Shen}, {Mulchaey}, {Raychaudhury},
  {Rasmussen}  \& {Ponman}}{{Shen} et~al.}{2007}]{Shen_2007}
{Shen} Y.,  {Mulchaey} J.~S.,  {Raychaudhury} S.,  {Rasmussen} J.,   {Ponman}
  T.~J.,  2007, \mn@doi [\apjl] {10.1086/511030}, 654, L115

\bibitem[\protect\citeauthoryear{{Shen} et~al.,}{{Shen}
  et~al.}{2009}]{Shen_2009}
{Shen} Y.,  et~al., 2009, \mn@doi [\apj] {10.1088/0004-637X/697/2/1656}, 697,
  1656

\bibitem[\protect\citeauthoryear{{Silverman} et~al.,}{{Silverman}
  et~al.}{2008}]{Silverman_2008}
{Silverman} J.~D.,  et~al., 2008, \mn@doi [\apj] {10.1086/529572}, 679, 118

\bibitem[\protect\citeauthoryear{{Simpson}, {Westoby}, {Arumugam}, {Ivison},
  {Hartley}  \& {Almaini}}{{Simpson} et~al.}{2013}]{Simpson_2013}
{Simpson} C.,  {Westoby} P.,  {Arumugam} V.,  {Ivison} R.,  {Hartley} W.,
  {Almaini} O.,  2013, \mn@doi [\mnras] {10.1093/mnras/stt940}, 433, 2647

\bibitem[\protect\citeauthoryear{{Smith} et~al.,}{{Smith}
  et~al.}{2003}]{Smith_2003}
{Smith} R.~E.,  et~al., 2003, \mn@doi [\mnras]
  {10.1046/j.1365-8711.2003.06503.x}, 341, 1311

\bibitem[\protect\citeauthoryear{{Stanley}, {Harrison}, {Alexander},
  {Swinbank}, {Aird}, {Del Moro}, {Hickox}  \& {Mullaney}}{{Stanley}
  et~al.}{2015}]{Stanley_2015}
{Stanley} F.,  {Harrison} C.~M.,  {Alexander} D.~M.,  {Swinbank} A.~M.,  {Aird}
  J.~A.,  {Del Moro} A.,  {Hickox} R.~C.,   {Mullaney} J.~R.,  2015, \mn@doi
  [\mnras] {10.1093/mnras/stv1678}, 453, 591

\bibitem[\protect\citeauthoryear{{Sutherland} \& {Saunders}}{{Sutherland} \&
  {Saunders}}{1992}]{Sutherland_Saunders_1992}
{Sutherland} W.,  {Saunders} W.,  1992, \mn@doi [\mnras]
  {10.1093/mnras/259.3.413}, 259, 413

\bibitem[\protect\citeauthoryear{{Tzanavaris} et~al.,}{{Tzanavaris}
  et~al.}{2014}]{Tzanavaris_2014}
{Tzanavaris} P.,  et~al., 2014, \mn@doi [\apjs] {10.1088/0067-0049/212/1/9},
  212, 9

\bibitem[\protect\citeauthoryear{{Ueda} et~al.,}{{Ueda}
  et~al.}{2008}]{Ueda_2008}
{Ueda} Y.,  et~al., 2008, \mn@doi [\apjs] {10.1086/591083}, 179, 124

\bibitem[\protect\citeauthoryear{{Villforth} et~al.,}{{Villforth}
  et~al.}{2017}]{Villforth_2017}
{Villforth} C.,  et~al., 2017, \mn@doi [\mnras] {10.1093/mnras/stw3037}, 466,
  812

\bibitem[\protect\citeauthoryear{{Wake} et~al.,}{{Wake}
  et~al.}{2011}]{Wake_2011}
{Wake} D.~A.,  et~al., 2011, \mn@doi [\apj] {10.1088/0004-637X/728/1/46}, 728,
  46

\bibitem[\protect\citeauthoryear{{Wild}, {Heckman}  \& {Charlot}}{{Wild}
  et~al.}{2010}]{Wild_2010}
{Wild} V.,  {Heckman} T.,   {Charlot} S.,  2010, \mn@doi [\mnras]
  {10.1111/j.1365-2966.2010.16536.x}, 405, 933

\bibitem[\protect\citeauthoryear{{Wild} et~al.,}{{Wild}
  et~al.}{2014}]{Wild_2014}
{Wild} V.,  et~al., 2014, \mn@doi [\mnras] {10.1093/mnras/stu212}, 440, 1880

\bibitem[\protect\citeauthoryear{Wild, Almaini, Dunlop, Simpson, Rowlands,
  Bowler, Maltby  \& McLure}{Wild et~al.}{2016}]{Wild_2016}
Wild V.,  Almaini O.,  Dunlop J.,  Simpson C.,  Rowlands K.,  Bowler R.,
  Maltby D.,   McLure R.,  2016, \mn@doi [\mnras] {10.1093/mnras/stw1996}, 463,
  832

\bibitem[\protect\citeauthoryear{{Wilkinson} et~al.,}{{Wilkinson}
  et~al.}{2017}]{Wilkinson_2017}
{Wilkinson} A.,  et~al., 2017, \mn@doi [\mnras] {10.1093/mnras/stw2405}, 464,
  1380

\bibitem[\protect\citeauthoryear{{Yang}, {Brandt}, {Darvish}, {Chen}, {Vito},
  {Alexander}, {Bauer}  \& {Trump}}{{Yang} et~al.}{2018}]{Yang_2018}
{Yang} G.,  {Brandt} W.~N.,  {Darvish} B.,  {Chen} C.-T.~J.,  {Vito} F.,
  {Alexander} D.~M.,  {Bauer} F.~E.,   {Trump} J.~R.,  2018, \mn@doi [\mnras]
  {10.1093/mnras/sty1910}, 480, 1022

\bibitem[\protect\citeauthoryear{{Zehavi} et~al.,}{{Zehavi}
  et~al.}{2005}]{Zehavi_2005}
{Zehavi} I.,  et~al., 2005, \mn@doi [\apj] {10.1086/431891}, 630, 1

\makeatother
\end{thebibliography}
\bsp


\label{lastpage}

\end{document}